\documentclass[12pt,twoside]{amsart}
\usepackage{amssymb}
\usepackage[mathscr]{eucal}
\usepackage{epic,eepic}

\newtheorem{prop}{Proposition}
\newcommand{\ds}{\displaystyle}
\def\EXP{\mbox{{\large\bf e}}}
\newcommand{\uop}{{u}}
\newcommand{\wop}{{w}}

\newcommand{\lsp}{\lambda}
\newcommand{\msp}{\mu}
\def\pot{\bi{n}}
\newcommand{\one}{\bi{e}_1}
\newcommand{\two}{\bi{e}_2}
\newcommand{\thr}{\bi{e}_3}
\def\r#1{(\ref{#1})}
\def\sk#1{\scriptstyle{#1}}
\let\bi=\mathbf
\let\ead=\email
\def\ftn#1{\footnotesize{#1}}
\renewcommand{\Box}{\square}

\begin{document}

\title[Discrete integrable 3-dim model]
{Spectral curves and parameterizations of\\
 a discrete integrable 3-dimensional model}
\author[S. Pakuliak and S. Sergeev]
{S. Pakuliak\dag\ and S. Sergeev\dag\ddag}
\address{\dag\ Bogoliubov
Laboratory of Theoretical Physics, Joint Institute
for Nuclear Research, Dubna 141980, Moscow reg., Russia}
\address{\ddag\ Max Plank Institute of Mathematics,
Vivatsgasse 7, D-53111, Bonn, Germany}
\ead{sergeev@mpim-bonn.mpg.de, pakuliak@thsun1.jinr.ru}
\begin{abstract}
We consider a discrete classical integrable model on the
3-dimensional cubic
lattice. The solutions of this model can be used to
parameterize the Boltzmann weights of the different 3-dimensional
spin models. We have found
the general solution of this model constructed in
terms of the theta-functions defined on an arbitrary compact
algebraic curve. The imposing of the periodic boundary
conditions fixes the algebraic curve. We have shown that in this
case the curve coincides with the spectral one of the auxiliary
linear problem. In the case when the curve is a rational one, the
soliton solutions have been constructed.
\end{abstract}

\maketitle


\section{Introduction}

The  paper is devoted to a description of the periodic and soliton
solutions of some generic  classical 3-dimensional discrete
integrable model. We will see that this description can be
presented in the completely analogous way as the description of
the finite-gap solutions of the hierarchy of the continuous
integrable equations, although the way we get the system of
equation is unusual and motivated by the approach developed in
\cite{Sergeev} for discrete integrable spin models. First, we
construct the discrete equations of motion from an equivalence of
the linear systems (which replace zero-curvature condition for Lax
operators in usual formulation) and  after that prove the
integrability by the counting the independent number of the
integrals of motion.

Several types of the boundary conditions may be considered on the
cubic lattice (the open boundary, periodical boundary conditions
in chosen directions, or the completely periodical boundary
conditions). This choice leads to several dynamical interpretation
of the model: as a Cauchy problem, as a B\"acklund transformation,
or as an analogue of the standing vibrations on the discrete $3d$
torus.

Starting from the discrete equations of motion we introduce a
change of variables through triple of the Legendre variables,
which transforms the equations of motion into the three-linear
form. These three-linear equations appear to be a generalization
of the famous Hirota bi-linear discrete equation. Then we observe
that these three-linear equations
can be formally solved with the help of Fay's identity for a
theta-function on an arbitrary algebraic curve.

Some facts observed in this paper are the demonstration of the
general statements proved more than two decades  ago in
\cite{Kr78} -- any discrete integrable system can be solved using
the algebraic-geometry methods. In this paper we develop further
an alternative approach to 3-dimensional discrete integrable
systems \cite{Sergeev} which does not use the notion of Lax
operators and can be applied equivalently to quantum (spin) or
classical integrable systems associated with several 3-dimensional
lattices. Instead of the Lax operators we use the notion of the
linear system defined on auxiliary planes. In this paper we
consider the cubic lattice, although these methods may be applied
to arbitrary 3-dimensional lattices formed by a set of planes.

\section{Classical discrete integrable system on the cubic lattice}

Let the vertices of the cubic lattice  spanned by orthogonal basis
$\one$, $\two$ and $\thr$ be marked by the vector
\begin{equation}\label{pot}
\pot=n_1\one+ n_2\two + n_3\thr\ .
\end{equation}

\begin{figure}
\setlength{\unitlength}{0.0004in}
\begin{center}
{\renewcommand{\dashlinestretch}{30}
\begin{picture}(8027,2964)(0,-10)
\path(237,2262)(6087,2262)
\blacken\path(5967.000,2232.000)(6087.000,2262.000)
(5967.000,2292.000)(5967.000,2232.000)
\path(237,1587)(6087,1587)
\blacken\path(5967.000,1557.000)(6087.000,1587.000)
(5967.000,1617.000)(5967.000,1557.000)
\path(237,912)(6087,912)
\blacken\path(5967.000,882.000)(6087.000,912.000)
(5967.000,942.000)(5967.000,882.000)
\path(462,1812)(1137,2487)
\blacken\path(1073.360,2380.934)(1137.000,2487.000)
(1030.934,2423.360)(1073.360,2380.934)
\path(462,1137)(1137,1812)
\blacken\path(1073.360,1705.934)(1137.000,1812.000)
(1030.934,1748.360)(1073.360,1705.934)
\path(462,462)(1137,1137)
\blacken\path(1073.360,1030.934)(1137.000,1137.000)
(1030.934,1073.360)(1073.360,1030.934)
\path(12,2037)(5862,2037)
\blacken\path(5742.000,2007.000)(5862.000,2037.000)
(5742.000,2067.000)(5742.000,2007.000)
\path(12,1362)(5862,1362)
\blacken\path(5742.000,1332.000)(5862.000,1362.000)
(5742.000,1392.000)(5742.000,1332.000)
\path(12,687)(5862,687)
\blacken\path(5742.000,657.000)(5862.000,687.000)
(5742.000,717.000)(5742.000,657.000)
\path(2037,12)(2037,2487)
\blacken\path(2067.000,2367.000)(2037.000,2487.000)
(2007.000,2367.000)(2067.000,2367.000)
\path(2262,462)(2262,2937)
\blacken\path(2292.000,2817.000)(2262.000,2937.000)
(2232.000,2817.000)(2292.000,2817.000)
\path(1812,462)(2487,1137)
\blacken\path(2423.360,1030.934)(2487.000,1137.000)
(2380.934,1073.360)(2423.360,1030.934)
\path(1812,1137)(2487,1812)
\blacken\path(2423.360,1705.934)(2487.000,1812.000)
(2380.934,1748.360)(2423.360,1705.934)
\path(1812,1812)(2487,2487)
\blacken\path(2423.360,2380.934)(2487.000,2487.000)
(2380.934,2423.360)(2423.360,2380.934)
\path(687,12)(687,2487)
\blacken\path(717.000,2367.000)(687.000,2487.000)
(657.000,2367.000)(717.000,2367.000)
\path(912,462)(912,2937)
\blacken\path(942.000,2817.000)(912.000,2937.000)
(882.000,2817.000)(942.000,2817.000)
\path(3387,12)(3387,2487)
\blacken\path(3417.000,2367.000)(3387.000,2487.000)
(3357.000,2367.000)(3417.000,2367.000)
\path(3612,462)(3612,2937)
\blacken\path(3642.000,2817.000)(3612.000,2937.000)
(3582.000,2817.000)(3642.000,2817.000)
\path(3162,1812)(3837,2487)
\blacken\path(3773.360,2380.934)(3837.000,2487.000)
(3730.934,2423.360)(3773.360,2380.934)
\path(3162,1137)(3837,1812)
\blacken\path(3773.360,1705.934)(3837.000,1812.000)
(3730.934,1748.360)(3773.360,1705.934)
\path(3162,462)(3837,1137)
\blacken\path(3773.360,1030.934)(3837.000,1137.000)
(3730.934,1073.360)(3773.360,1030.934)
\path(4737,12)(4737,2487)
\blacken\path(4767.000,2367.000)(4737.000,2487.000)
(4707.000,2367.000)(4767.000,2367.000)
\path(4962,462)(4962,2937)
\blacken\path(4992.000,2817.000)(4962.000,2937.000)
(4932.000,2817.000)(4992.000,2817.000)
\path(4512,1812)(5187,2487)
\blacken\path(5123.360,2380.934)(5187.000,2487.000)
(5080.934,2423.360)(5123.360,2380.934)
\path(4512,1137)(5187,1812)
\blacken\path(5123.360,1705.934)(5187.000,1812.000)
(5080.934,1748.360)(5123.360,1705.934)
\path(4512,462)(5187,1137)
\blacken\path(5123.360,1030.934)(5187.000,1137.000)
(5080.934,1073.360)(5123.360,1030.934)
\end{picture}
}
\caption{\label{cubic}\ftn{Example of cubic lattice of the size
$3\times2\times4$.} }
\end{center}
\end{figure}
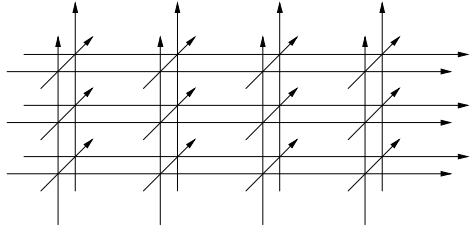

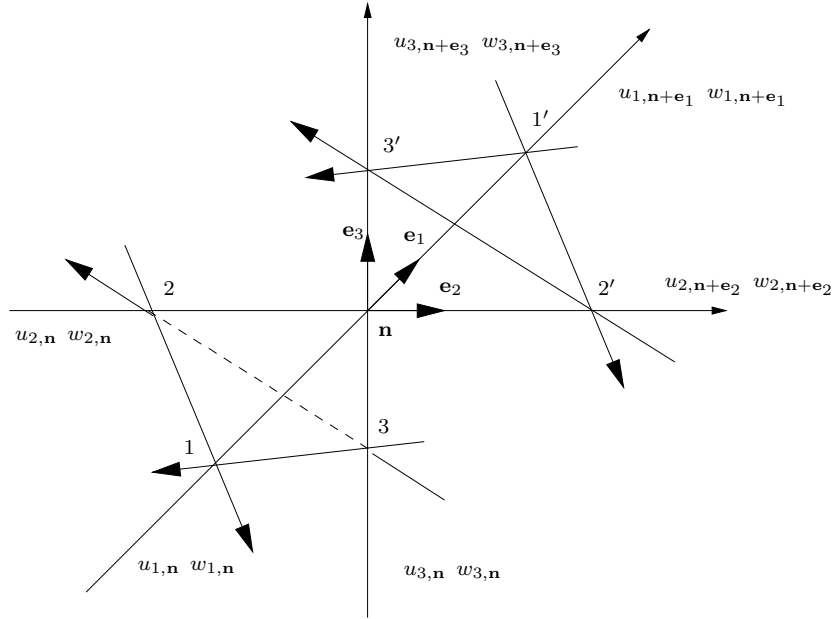
\begin{figure}
\setlength{\unitlength}{0.0006in}
\begin{center}
{\renewcommand{\dashlinestretch}{30}
\begin{picture}(6324,5439)(0,-10)
\path(3162,12)(3162,5412)
\blacken\path(3192.000,5292.000)(3162.000,5412.000)
(3132.000,5292.000)(3192.000,5292.000)
\path(12,2712)(6312,2712)
\blacken\path(6192.000,2682.000)(6312.000,2712.000)
(6192.000,2742.000)(6192.000,2682.000)
\path(687,237)(5637,5187)
\blacken\path(5573.360,5080.934)(5637.000,5187.000)
(5530.934,5123.360)(5573.360,5080.934)
\path(4287,4737)(5412,2037)
\blacken\path(5264.308,2235.462)(5412.000,2037.000)
(5375.077,2281.615)(5264.308,2235.462)
\path(5862,2262)(2487,4377)
\blacken\path(2722.228,4300.398)(2487.000,4377.000)
(2658.506,4198.715)(2722.228,4300.398)
\path(1025,3288)(2150,588)
\blacken\path(2002.308,786.462)(2150.000,588.000)
(2113.077,832.615)(2002.308,786.462)
\path(3655,1559)(1270,1289)
\blacken\path(1501.727,1375.617)(1270.000,1289.000)
(1515.226,1256.378)(1501.727,1375.617)
\path(5007,4154)(2622,3884)
\blacken\path(2853.727,3970.617)(2622.000,3884.000)
(2867.226,3851.378)(2853.727,3970.617)
\path(3162,2712)(3162,3387)
\blacken\path(3222.000,3147.000)(3162.000,3387.000)
(3102.000,3147.000)(3222.000,3147.000)
\path(3162,2712)(3837,2712)
\blacken\path(3597.000,2652.000)(3837.000,2712.000)
(3597.000,2772.000)(3597.000,2652.000)
\path(3162,2712)(3612,3162)
\blacken\path(3484.721,2949.868)(3612.000,3162.000)
(3399.868,3034.721)(3484.721,2949.868)
\dashline{90.000}(3162,1497)(1227,2712)
\path(3207,1452)(3837,1047) \path(1272,2667)(507,3162)
\blacken\path(741.092,3081.994)(507.000,3162.000)
(675.902,2981.245)(741.092,3081.994)
\put(3252,2487){\makebox(0,0)[lb]{$\sk{\pot}$}}
\put(3792,2847){\makebox(0,0)[lb]{$\sk{\two}$}}
\put(3477,3297){\makebox(0,0)[lb]{$\sk{\one}$}}
\put(2937,3342){\makebox(0,0)[lb]{$\sk{\thr}$}}
\put(3477,327){\makebox(0,0)[lb]
{$\sk{\uop_{3,\pot}\ \wop_{3,\pot}}$}}
\put(1137,372){\makebox(0,0)[lb]
{$\sk{\uop_{1,\pot}\ \wop_{1,\pot}}$}}
\put(57,2397){\makebox(0,0)[lb]
{$\sk{\uop_{2,\pot}\ \wop_{2,\pot}}$}}
\put(5772,2847){\makebox(0,0)[lb]
{$\sk{\uop_{2,\pot+\two}\ \wop_{2,\pot+\two}}$}}
\put(5367,4512){\makebox(0,0)[lb]
{$\sk{\uop_{1,\pot+\one}\ \wop_{1,\pot+\one}}$}}
\put(3387,4962){\makebox(0,0)[lb]
{$\sk{\uop_{3,\pot+\thr}\ \wop_{3,\pot+\thr}}$}}
\put(1542,1452){\makebox(0,0)[lb]{$\sk{1}$}}
\put(4602,4332){\makebox(0,0)[lb]{$\sk{1'}$}}
\put(3252,1632){\makebox(0,0)[lb]{$\sk{3}$}}
\put(3297,4107){\makebox(0,0)[lb]{$\sk{3'}$}}
\put(5187,2847){\makebox(0,0)[lb]{$\sk{2'}$}}
\put(1362,2847){\makebox(0,0)[lb]{$\sk{2}$}}
\end{picture}
}
\caption{\label{single-vertex}
\ftn{Association of the dynamical variables
to the edges of the cubic lattice. Two intersecting planes form
two triangles $123$ and $1'2'3'$.}}
\end{center}
\end{figure}

The cubic lattice is formed by three sets of parallel planes
or, equivalently, by three sets of parallel lines (see
Fig.~\ref{cubic}). Let us associate to each edge of the cubic
lattice of a given type $\alpha$, ($\alpha=1,2,3$ corresponds to
three orthogonal directions of the lattice) the dynamical
variables as shown in  Fig.~\ref{single-vertex}. Namely, the
pairs of dynamical variables $\uop_{\alpha,\pot},
\wop_{\alpha,\pot}$, $\alpha=1,2,3$, are associated with the edges
incoming to the oriented vertex $\pot$, while
$\uop_{\alpha,\pot+\bf{e}_{\alpha}},
\wop_{\alpha,\pot+\bf{e}_{\alpha}}$ are associated to the outgoing
edges. Fig.~\ref{single-vertex} shows also two auxiliary
planes intersecting the incoming and outgoing edges. Each of these
planes cuts seven of eight octants around the vertex of the cubic
lattice numbered by the vector $\pot$.

Let us stress that our considerations at the moment are local. Our
goal is to obtain the relations between dynamical variables
surrounding given vertex. After this we extend these relations to
the whole lattice and so obtain the discrete dynamical system.

\begin{figure}
\setlength{\unitlength}{0.0004in}
\begin{center}
{\renewcommand{\dashlinestretch}{30}
\begin{picture}(3644,3659)(0,-10)
\thicklines
\path(1822,22)(1822,3622)
\blacken\path(1882.000,3382.000)
(1822.000,3622.000)(1762.000,3382.000)(1882.000,3382.000)
\path(22,1822)(3622,1822)
\blacken\path(3382.000,1762.000)
(3622.000,1822.000)(3382.000,1882.000)(3382.000,1762.000)
\put(557,2930){\makebox(0,0)[lb]{$\Phi_a$}}
\put(3000,2930){\makebox(0,0)[lb]{$\Phi_b$}}
\put(547,680){\makebox(0,0)[lb]{$\Phi_c$}}
\put(3000,500){\makebox(0,0)[lb]{$\Phi_d$}}
\put(2075,1175){\makebox(0,0)[lb]{$\uop,\wop,\kappa$}}
\thinlines
\path(2047,1620)(2047,1170)(3372,1170)
    (3372,1620)(2047,1620)(1822,1825)
\blacken\path(1958.066,1751.360)(1862.000,1785.000)
(1915.640,1668.934)(1958.066,1751.360)
\put(1725,1725){\makebox(0,0)[lb]{$\bullet$}}
\end{picture}
}
\end{center}
\caption{\label{vertex}\ftn{The linear problem for the  vertex
with associated dynamical
variables and additional parameter $\kappa$.}}
\end{figure}
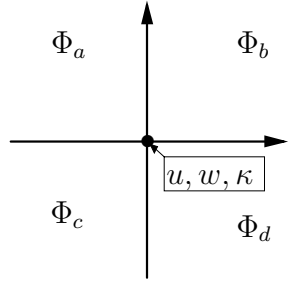

Each of the triangles on the auxiliary planes is formed by three
lines obtained by the intersection of this plane with planes
forming the vertex $\pot$ of the cubic lattice. In this way the
vertices of the auxiliary triangles are associated to the edges of
the cubic lattice and so to a pair of the dynamical variables. Let
us consider two linear problems attached to the auxiliary
triangles via the following rules. First, we introduce the linear
variables $\Phi_a$, $\Phi_b$, $\Phi_c$ and $\Phi_d$ around the
vertex on the auxiliary plane according to  Fig.~\ref{vertex}.
The linear problems for each  vertex on the auxiliary planes will
be always of the form
\begin{equation}\label{lin-pr}
0=\Phi_a-\Phi_b\cdot\uop+\Phi_c\cdot\wop+\Phi_d\cdot\kappa\;
\uop\;\wop\;.
\end{equation}
Here $\kappa\in\mathbb{C}$ is an additional parameter associated
with the line of the cubic lattice. Note that the coefficients of
the linear form \r{lin-pr} are fixed by the orientation of the
lines which form the vertex on the auxiliary plane.

\begin{figure}
\setlength{\unitlength}{0.0006in}
\begin{center}
{\renewcommand{\dashlinestretch}{30}
\begin{picture}(8349,3874)(0,-10)
\path(12,687)(4062,687)
\blacken\path(3942.000,657.000)(4062.000,687.000)
(3942.000,717.000)(3942.000,657.000) \path(687,12)(2937,3837)
\blacken\path(2902.016,3718.357)(2937.000,3837.000)
(2850.300,3748.778)(2902.016,3718.357) \path(3387,12)(1137,3837)
\blacken\path(1223.700,3748.778)(1137.000,3837.000)
(1171.984,3718.357)(1223.700,3748.778) \path(4287,3162)(8337,3162)
\blacken\path(8217.000,3132.000)(8337.000,3162.000)
(8217.000,3192.000)(8217.000,3132.000) \path(7185,22)(4935,3847)
\blacken\path(5021.700,3758.778)(4935.000,3847.000)
(4969.984,3728.357)(5021.700,3758.778) \path(5406,15)(7656,3840)
\blacken\path(7621.016,3721.357)(7656.000,3840.000)
(7569.300,3751.778)(7621.016,3721.357)
\put(3400,250){\makebox(0,0)[lb]
{$\Phi_{\mathbf{n}+\mathbf{e}_1+\mathbf{e}_3}$}}
\put(1700,250){\makebox(0,0)[lb]
{$\Phi_{\mathbf{n}+\mathbf{e}_1}$}}
\put(-300,250){\makebox(0,0)[lb]
{$\Phi_{\mathbf{n}+\mathbf{e}_1+\mathbf{e}_2}$}}
\put(700,2000){\makebox(0,0)[lb]
{$\Phi_{\mathbf{n}+\mathbf{e}_2}$}}
\put(1900,1300){\makebox(0,0)[lb]{$\Phi_{\mathbf{n}}$}}
\put(2500,2000){\makebox(0,0)[lb]
{$\Phi_{\mathbf{n}+\mathbf{e}_3}$}}
\put(1500,3400){\makebox(0,0)[lb]
{$\Phi_{\mathbf{n}+\mathbf{e}_2+\mathbf{e}_3}$}}
\put(5600,2800){\makebox(0,0)[lb]
{$\Phi_{\mathbf{n}+\mathbf{e}_1+\mathbf{e}_2+\mathbf{e}_3}$}}
\put(5700,3400){\makebox(0,0)[lb]
{$\Phi_{\mathbf{n}+\mathbf{e}_2+\mathbf{e}_3}$}}
\put(4100,3400){\makebox(0,0)[lb]
{$\Phi_{\mathbf{n}+\mathbf{e}_2}$}}
\put(7932,3400){\makebox(0,0)[lb]
{$\Phi_{\mathbf{n}+\mathbf{e}_3}$}}
\put(6000,400){\makebox(0,0)[lb]
{$\Phi_{\mathbf{n}+\mathbf{e}_1}$}}
\put(4900,1400){\makebox(0,0)[lb]
{$\Phi_{\mathbf{n}+\mathbf{e}_1+\mathbf{e}_2}$}}
\put(6700,1400){\makebox(0,0)[lb]
{$\Phi_{\mathbf{n}+\mathbf{e}_1+\mathbf{e}_3}$}}
\put(4000,2037){\makebox(0,0)[lb]{$\sim$}}
\end{picture}
} \caption{\label{mean-xi}\ftn{Parameterizations
of the classical linear variables
$\Phi_\pot$ in cubic geometry.}}
\end{center}
\end{figure}
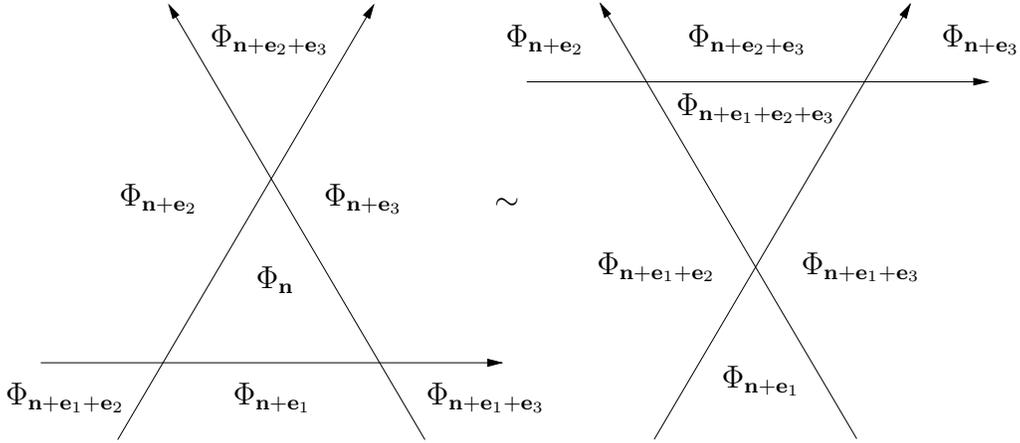
The enumeration of the linear variables on the auxiliary triangles
is shown in Fig.~\ref{mean-xi}. It is clear that these linear
variables are associated with internal part of the cubes in the
cubic lattice. According to $\r{lin-pr}$ we can write the system of
the linear equations
\begin{equation}\label{eq-in}
\left\{\begin{array}{lcr}
0&=&\ds\Phi_{\pot+\two}-\Phi_{\pot+\two+\thr}u_{1,\pot}+
\Phi_{\pot}w_{1,\pot}+\Phi_{\pot+\thr}\kappa_{1:\pot}u_{1,\pot}
w_{1,\pot}\;,\\
0&=&\ds\Phi_{\pot}-\Phi_{\pot+\thr}u_{2,\pot}+
\Phi_{\pot+\one}w_{2,\pot}+\Phi_{\pot+\one+\thr}
\kappa_{2:\pot}u_{2,\pot}
w_{2,\pot}\;,\\
0&=&\ds\Phi_{\pot+\two}-\Phi_{\pot}u_{3,\pot}+
\Phi_{\pot+\one+\two}w_{3,\pot}+\Phi_{\pot+\one}
\kappa_{3:\pot}u_{3,\pot}
w_{3,\pot}\;
\end{array}\right.
\end{equation}
for left triangle shown in Fig.~\ref{mean-xi} and the
linear system
\begin{equation}\label{eq-out}
\left\{
\begin{array}{rcl}
0&=&\ds\Phi_{\pot+\one+\two}-\Phi_{\pot+\one+\two+\thr}
u_{1,\pot+\one}+
\Phi_{n+\one} w_{1,\pot+\one}+\\
&&+\
\Phi_{\pot+\one+\thr}\kappa_{1,\pot}u_{1,\pot+\one}w_{1,\pot+\one}\;,\\
0&=&\ds\Phi_{\pot+\two}-\Phi_{\pot+\two+\thr}
u_{2,\pot+\two} +\Phi_{n+\one+\two} w_{2,\pot+\two}+\\
&&+\
\Phi_{\pot+\one+\two+\thr}\kappa_{2,\pot}u_{2,\pot+\two}
w_{2,\pot+\two}\;,\\
0&=&\Phi_{\pot+\two+\thr}-\Phi_{\pot+\thr}u_{3,\pot+\thr}
+\Phi_{n+\one+\two+\thr} w_{3,\pot+\thr}+\\
&&+\ \Phi_{\pot+\two+\thr}\kappa_{3,\pot}u_{3,\pot+\thr}
w_{3,\pot+\thr}
\end{array}
\right.
\end{equation}
for the right one.

Supposing that dynamical variables do not vanish identically on
any edge of the lattice we may exclude from the systems \r{eq-in}
and \r{eq-out} the linear variables $\Phi_\pot$ and
$\Phi_{\pot+\one+\two+\thr}$. Let us require that the systems
\r{eq-in} and \r{eq-out} (each contains now only two linear
relations for the rest 6 linear variables $\Phi_{\pot+\one}$,
$\Phi_{\pot+\one+\two}$, etc.) are equivalent. This means that the
matrix elements of these linear systems which are rational
functions of all dynamical variables around given vertex coincide
identically. Moreover, let us require also that each type of the
parameters $\kappa_{\alpha,\pot}$ is conserved along the
corresponding direction (which is already taken into account in
\r{eq-out})
\begin{equation}\label{conser}
\kappa_{\alpha,\pot}=\kappa_{\alpha,\pot+\bf{e}_\alpha}\ ,
\quad\alpha=1,2,3.
\end{equation}
As the result, we obtain the recurrent relations for dynamical
variables $\uop_{\alpha,\pot}$, $\wop_{\alpha,\pot}$,
$\uop_{\alpha,\pot+\bi{e}_\alpha}$ and
$\wop_{\alpha,\pot+\bi{e}_\alpha}$:
\begin{equation}\label{eq-of-mo1}
\left\{\begin{array}{l} \ds u_{1,\pot+\one} \;=\;
\frac{\kappa_{2,\pot}\,u_{1,\pot}\,u_{2,\pot}\,w_{2,\pot}}
{\kappa_{1,\pot}\,u_{1,\pot}\,w_{2,\pot} \,+\,
\kappa_{3,\pot}\,u_{2,\pot}\,w_{3,\pot} \,+\,
\kappa_{1,\pot}\,\kappa_{3,\pot}\,u_{1,\pot}\,w_{3,\pot}}\;,\\
\\
\ds w_{1,\pot+\one} \;=\; \frac{w_{1,\pot}\,w_{2,\pot} \,+\,
u_{3,\pot}\,w_{2,\pot} \,+\,
\kappa_{3,\pot}\,u_{3,\pot}\,w_{3,\pot}}{ w_{3,\pot} }\;,
\end{array}\right.
\end{equation}
\begin{equation}\label{eq-of-mo2}
\left\{\begin{array}{l} \ds u_{2,\pot+\two} \;=\;
\frac{u_{1,\pot}\,u_{2,\pot}\,u_{3,\pot}}{
u_{2,\pot}\,u_{3,\pot} \,+\, u_{2,\pot}\, w_{1,\pot}
\,+\, \kappa_{1,\pot}\,u_{1,\pot}\,w_{1,\pot} }\;,\\
\\
\ds w_{2,\pot+\two} \;=\;
\frac{w_{1,\pot}\,w_{2,\pot}\,w_{3,\pot} }{
w_{1,\pot}\,w_{2,\pot} \,+\, u_{3,\pot}\,w_{2,\pot} \,+\,
\kappa_{3,\pot}\,u_{3,\pot}\,w_{3,\pot} }\;,
\end{array}\right.
\end{equation}
\begin{equation}\label{eq-of-mo3}
\left\{\begin{array}{l} \ds u_{3,\pot+\thr} \;=\;
\frac{ u_{2,\pot}\,u_{3,\pot}  \,+\, u_{2,\pot}\, w_{1,\pot}
\,+\, \kappa_{1,\pot}\,u_{1,\pot}\,w_{1,\pot} }{
u_{1,\pot}}\;,\\
\\
\ds w_{3,\pot+\thr} \;=\;
\frac{\kappa_{2,\pot}\,u_{2,\pot}\,w_{2,\pot}\,w_{3,\pot} }{
\kappa_{1,\pot}\,u_{1,\pot}\,w_{2,\pot} \,+\,
\kappa_{3,\pot}\,u_{2,\pot}\,w_{3,\pot} \,+\,
\kappa_{1,\pot}\,\kappa_{3,\pot}\,u_{1,\pot}\,w_{3,\pot}}\;.
\end{array}\right.
\end{equation}
where due to \r{conser} we have
\begin{equation}\label{kappa-pref}
\ds \kappa_{1,\pot}\;=\;\kappa_{1:\,n_2,n_3}\;,\;\;\;
\kappa_{2,\pot}\;=\;\kappa_{2:\,n_1,n_3}\;,\;\;\;
\kappa_{3,\pot}\;=\;\kappa_{3:\,n_1,n_2}\;.
\end{equation}
One may ask why the recurrent relations \r{eq-of-mo1}--\r{eq-of-mo3}
are the dynamical system.
Let us introduce the following Poisson bracket for the edges
incoming to the vertex with some fixed $\pot$:
\begin{equation}\label{Poisson}
\{\uop_{\alpha,\pot},\wop_{\alpha,\pot}\}=\uop_{\alpha,\pot}
\wop_{\alpha,\pot}
\end{equation}
and any other bracket for incoming edges vanishes. The
transformation \r{eq-of-mo1}--\r{eq-of-mo3} is the canonical one,
i.e. from \r{Poisson} it follows that
\begin{equation}
\{ \uop_{\alpha,\pot+\bi{e}_\alpha},
\wop_{\alpha,\pot+\bi{e}_\alpha} \}=
\uop_{\alpha,\pot+\bi{e}_\alpha}
\wop_{\alpha,\pot+\bi{e}_\alpha}
\end{equation}
and any other bracket for outgoing edges vanishes. In general, the
auxiliary triangles shown in Fig. \r{single-vertex} are the
fragments of two adjacent space-like surfaces, and the set of 3d
vertices between these surfaces provides by means of
\r{eq-of-mo1}--\r{eq-of-mo3} the canonical transformation of the
set of dynamical pairs from incoming surface to the outgoing one.
We call this transformation the local equation of motion in the
direction perpendicular to the chosen space-like surface. Besides
the local equations of motion \r{eq-of-mo1}--\r{eq-of-mo3}, a
complete formulation of a dynamical system needs the specification
of boundary conditions.

The type of the boundary conditions as well as the global
characteristics of the model depend on the choice of the
space-like surface. The auxiliary lattices described are
appropriate for the Cauchy problem for the cubic lattice. In
Section~\ref{integrability} we will choose $n_1$ as the discrete
time, while $n_2$ and $n_3$ will be the discrete space
coordinates. At this choice the system \r{eq-of-mo1} will describe
the ``time'' evolution of the variables $\uop_{1,\pot}$,
$\wop_{1,\pot}$ and the systems \r{eq-of-mo2}--\r{eq-of-mo3} --
the space distribution of the ``auxiliary'' for this evolution
variables $\uop_{2,\pot}$, $\wop_{2,\pot}$, $\uop_{3,\pot}$ and
$\wop_{3,\pot}$. All the dynamical quantities, such as integrals
of motion, spectral curves etc., should be calculated in this case
for the system \r{eq-of-mo1}. This will be subject of the
Section~\ref{integrability}.

\section{Legendre transform}

It follows from (\ref{eq-of-mo1})--\r{eq-of-mo3}
\begin{equation}\label{simple}
\begin{array}{rcl}
\ds w_{1,\pot}
w_{2,\pot}&=&\ds w_{1,\pot+\one} w_{2,\pot+\two}\;,\quad
u_{2,\pot}u_{3,\pot}=u_{2,\pot+\two}u_{3,\pot+\thr}\;,\\
&&\\
\ds\frac{u_{1,\pot}}{w_{3,\pot}} &=&\ds
\frac{u_{1,\pot+\one}}{w_{3,\pot+\thr}}\ .
\end{array}
\end{equation}
Relations (\ref{simple}) provide in general the possibility of the
following change of the variables:
\begin{equation}\label{Legendre}
\ds\begin{array}{rclrcl}
\ds u_{1,\pot}&=&\ds u_{1:\,n_2,n_3}^{(0)}
\frac{\tau_{2,\pot}}{\tau_{2,\pot+\thr}}\;,\quad&\quad
\ds w_{1,\pot}&=&\ds w_{1:\,n_2,n_3}^{(0)}
\,\frac{\tau_{3,\pot+\two}}{\tau_{3,\pot}}\;,\\
&&&&&\\
\ds u_{2,\pot}&=&\ds u_{2:\,n_1,n_3}^{(0)}\,
\frac{\tau_{1,\pot}}{\tau_{1,\pot+\thr}}\;,\quad&\quad
\ds w_{2,\pot}&=&\ds w_{2:\,n_1,n_3}^{(0)}
\,\frac{\tau_{3,\pot}}{\tau_{3,\pot+\one}}\;,\\
&&&&&\\
\ds u_{3,\pot}&=&\ds u_{3:\,n_1,n_2}^{(0)}\,
\frac{\tau_{1,\pot+\two}}{\tau_{1,\pot}}\;,\quad&\quad
\ds w_{3,\pot}&=&\ds w_{3:\,n_1,n_2}^{(0)}\,
\frac{\tau_{2,\pot}}{\tau_{2,\pot+\one}}\;.
\end{array}
\end{equation}
Eqs. (\ref{eq-of-mo1})--\r{eq-of-mo3} with
the substitution (\ref{Legendre}) become
\begin{equation}\label{ttt}
\ds\begin{array}{l} \ds r_{\alpha,\pot}
\tau_{\alpha,\pot+\mathbf{e}_\beta+\mathbf{e}_\gamma}
\tau_{\beta,\pot} \tau_{\gamma,\pot} =
\tau_{\alpha,\pot} \tau_{\beta,\pot+\mathbf{e}_\gamma}
\tau_{\gamma,\pot+\mathbf{e}_\beta}\; +\\
\quad +\;  s_{\beta,\pot}
\tau_{\alpha,\pot+\mathbf{e}_\beta}
\tau_{\beta,\pot+\mathbf{e}_\gamma} \tau_{\gamma,\pot} +
s_{\gamma,\pot}^{-1} \tau_{\alpha,\pot+\mathbf{e}_\gamma}
\tau_{\beta,\pot} \tau_{\gamma,\pot+\mathbf{e}_\beta}\;,
\end{array}
\end{equation}
where $(\alpha,\beta,\gamma)$ is any \emph{cyclic permutation} of
the indices $(1,2,3)$, and the coefficients in \r{ttt} are
\begin{equation}\label{rs-uwk}
\begin{array}{ll}
\ds s_{1,\pot} \;=\; \frac{\kappa_{3:\,n_1,n_2}^{}\,
w_{3:\,n_1,n_2}^{(0)}}{w_{2:\,n_1,n_3}^{(0)}}\;,& \ds r_{1,\pot}
\;=\; \frac{u_{1:\,n_2,n_3}^{(0)}\, u_{3:\,n_1,n_2}^{(0)}}{
w_{1:\,n_2,n_3}^{(0)}\, u_{2:\,n_1,n_3}^{(0)}}\,,\\
&\\
\ds s_{2,\pot} \;=\; \frac{u_{3:\,n_1,n_2}^{(0)}}{
w_{1:\,n_2,n_3}^{(0)}}\,,& \ds r_{2,\pot}\;=\;
\frac{\kappa_{2:\,n_1,n_3}^{}\, u_{2:\,n_1,n_3}^{(0)}\,
w_{2:\,n_1,n_3}^{(0)}}{\kappa_{1:\,n_2,n_3}^{}\,
\kappa_{3:\,n_1,n_2}^{} \, u_{1:\,n_2,n_3}^{(0)} \, w_{3:\,n_1,n_2}^{(0)}}\;,\\
&\\
\ds s_{3,\pot} \;=\; \frac{u_{2:\,n_1,n_3}^{(0)}}{
\kappa_{1:\,n_2,n_3}^{}\, u_{1:\,n_2,n_3}^{(0)}}\,,& \ds
r_{3,\pot}^{} \;=\; \frac{w_{1:\,n_2,n_3}^{(0)}\,
w_{3:\,n_1,n_2}^{(0)}}{w_{2:\,n_1,n_3}^{(0)} \,
u_{3:\,n_1,n_2}^{(0)}}\;.
\end{array}
\end{equation}
This change of variables, $\tau_{\alpha,\pot}$ instead of
$u_{\alpha,\pot}$ and $w_{\alpha,\pot}$, has the following
interpretation. Equations (\ref{eq-of-mo1})--\r{eq-of-mo3} are the
kind of the Hamiltonian equations of motion for the classical
discrete system. Substitution (\ref{Legendre}) therefore is the
Legendre transformation. And finally equations (\ref{ttt}) are the
Lagrangian equations of motion. We will call
$\tau_{\alpha,\pot}$ as the Legendre variables, and the
coefficients $u_{\alpha:\,n_\beta,n_\gamma}^{(0)}$ and
$w_{\alpha:\,n_\beta,n_\gamma}^{(0)}$ as the pre-exponents.

Equations \r{ttt} generalize in some sense the famous bi-linear
Hirota equation \cite{Hirota}. It can be obtained from \r{ttt} in
the special limit when all parameters
$\kappa_{\alpha:n_{\beta},n_{\gamma}}$ tends to zero. Indeed, let
the parameters $r_{2,\pot}$, $s_{3,\pot}$ and $s^{-1}_{1,\pot}$
which contains $\kappa$-s tend to infinity
\begin{equation}\label{tends}
r_{2,\pot}\sim \varepsilon^{-2}+\frac{1}{2}\varepsilon^{-1}\;,\quad
s_{3,\pot}\sim \varepsilon^{-2}-\frac{1}{2}\varepsilon^{-1}\;,\quad
s^{-1}_{1,\pot}\sim \varepsilon^{-1}
\end{equation}
when $\varepsilon\to0$. Equating in \r{ttt} with $\alpha=2$,
$\beta=3$ and $\gamma=1$ the coefficients at $\varepsilon^{-2}$
and $\varepsilon^{-1}$ we obtain respectively
\begin{equation}\label{identif}
\tau_{3,\pot}=\tau_{2,\pot+\thr}\;,\quad
\tau_{1,\pot}=\tau_{2,\pot+\one}\;.
\end{equation}
Note that in the limit \r{tends} $\kappa_{1,\pot} \sim
\varepsilon^2, \quad \kappa_{2,\pot} \sim \kappa_{3,\pot} \sim
\varepsilon$. Imposing further one more relation $r_{1,\pot} =
s_{2,\pot} r_{3,\pot}$, one can see that the rest two equations of
\r{ttt} coincide and can be written as a single difference
equation for the Legendre variable $\tau_{2,\pot}$
$$
r_{1,\pot}\tau_{2,\pot+\one+\two+\thr}\tau_{2,\pot}=
\tau_{2,\pot+\one}\tau_{2,\pot+\two+\thr}+s_{2,\pot}
\tau_{2,\pot+\thr}\tau_{2,\pot+\one+\two}\ .
$$
In the homogeneous limit for the parameters $r_{1,\pot}=r_1$ and
$s_{2,\pot}=s_2$ the latter equation can be rewritten in the
canonical Hirota's form after obvious re-enumeration of discrete
variables. Due to this analogy we will call sometimes the
Legendre variables $\tau_{j,\pot}$ as the triplet of
$\tau$-functions.

We conclude this section with a description of the discrete gauge
invariance of the linear systems (\ref{eq-in}) and (\ref{eq-out}).
Let us require that these systems are invariant under simultaneous
shift of the linear variables
\begin{equation}\label{gauge}
\Phi_{\pot}\mapsto\xi_{\pot}\Phi_{\pot}\ ,
\end{equation}
where  the set of $\xi_{\pot}\in\,\mathbb{C}$. The invariance
requires the corresponding change of dynamical variables
$$
u_{1,\pot}\mapsto
\frac{\xi_{\pot+\two}}{\xi_{\pot+\two+\thr}}u_{1,\pot}\;, \quad
w_{1,\pot}\mapsto
\frac{\xi_{\pot+\two}}{\xi_{\pot}}w_{1,\pot}\;,\quad
\mbox{etc.}
$$
Now we have to request that gauge transformation of the parameters
$\kappa_{\alpha,\pot}$, $\alpha=1,2,3$ in both linear systems
(\ref{eq-in}) and (\ref{eq-out}), coincide:
\begin{equation}\label{kappa-gauge}
\kappa_{1,\pot}\mapsto \frac{\xi_{\pot}\xi_{\pot+\two+\thr}}
{\xi_{\pot+\two}\xi_{\pot+\thr}} \kappa_{1,\pot}=
\frac{\xi_{\pot+\one}\xi_{\pot+\one+\two+\thr}}{\xi_{\pot+\one+\two}
\xi_{\pot+\one+\thr}}\kappa_{1,\pot}\;,
\end{equation}
and similarly for the parameters $\kappa_{2,\pot}$ and
$\kappa_{3,\pot}$. Any of this requirements leads the single
restriction
\begin{equation}\label{restric}
\xi_{\pot} \xi_{\pot+\one+\two} \xi_{\pot+\one+\thr}
\xi_{\pot+\two+\thr}= \xi_{\pot+\one} \xi_{\pot+\two} \xi_{\pot+\thr}
\xi_{\pot+\one+\two+\thr}\;.
\end{equation}
The last equation can be easily solved by the ansatz
\begin{equation}\label{ansatz}
\xi_{\pot}=\xi_{1:n_2,n_3}\xi_{2:n_1,n_3}\xi_{3:n_1,n_2}
\end{equation}
and explains appearing of all $\xi$-dependent factors in the
statement of the following Proposition~\ref{prop5}.

\section{General solution of the classical equations of motion}

As we have seen, equations (\ref{ttt}) generalize the famous
Hirota equation. Actually the structure of (\ref{ttt}) is the same
as the structure of the Hirota equation. In this section we
present a general solution of these classical equation of motion
when each three-linear equation \r{ttt} is reduced to a pair of
bi-linear equations. This form is suitable for imposing the
periodic boundary condition. The bi-linear relations become a
celebrated Fay's identities for the $\Theta$-functions associated
with an algebraic curve of a finite genus. Periodic boundary
condition will fix the generic algebraic curve in an unique way.
The curve may depend on the type of the boundary conditions. We
claim that for the periodic boundary conditions the found solution
of the three-linear equations \r{ttt} is the most general. We
start with formulation of the necessary algebraic geometry objects
(see e.g. \cite{Mumford,Fay}).

Let $\Gamma$ be an arbitrary algebraic curve of genus $g$, and
$\omega=(\omega_1,\omega_2,...\omega_g)$ be the vector of $g$
holomorphic differentials, normalized as usual:
\begin{equation}
\ds \oint_{a_j}\;\omega_k\;=\;\delta_{k,j}\;,\;\;\;
\oint_{b_j}\omega_k\;=\;\Omega_{k,j}\;,
\end{equation}
where $a_j$, $b_j$, $j=1,\ldots,g$ are the sets of canonical
cycles on $\Gamma$.

Let $\bi{I}$: $\Gamma^{\otimes 2}\mapsto\textrm{Jac}(\Gamma)$ be
the Jacobi transform
\begin{equation}
\ds X,Y\;\in\;\Gamma\;\mapsto
\bi{I}(X,Y)\;=\;\int_X^Y\;\bi{\omega}\;\in\textrm{Jac}(\Gamma)\;.
\end{equation}
Let further $\Theta_{\bi{\epsilon}} (\bi{v})$,
$\bi{v}\in\mathbb{C}^g$, $\epsilon=(\epsilon_1, \epsilon_2)$,
$\bi{\epsilon}_i\in\mathbb{C}^g$ be the theta-function with
characteristic $\epsilon$ on the Jacobian Jac$(\Gamma)$,
\begin{equation}\label{Theta}
\ds \Theta_{\bi{\epsilon}}(\bi{v})=\sum_{\bi{m}\in\,\mathbb{Z}^g}\;
\exp\left({i\pi(\bi{m}+\bi{\epsilon}_1,\Omega\bi{m}+\bi{\epsilon}_1)+
2i \pi (\bi{m}+\bi{\epsilon}_1,\bi{v}+\bi{\epsilon}_2
)}\right),
\end{equation}
and $E(X,Y)=-E(Y,X)$ be the prime form $X,Y\in\Gamma$, so that the
cross-ratio
\begin{equation}
\ds \frac{E(X,Y)\,E(X',Y')}{E(X,Y')\, E(X',Y)}=
\frac{\Theta_{\bi{\epsilon}_{\rm odd}}
(\bi{I}(X,Y))\Theta_{\bi{\epsilon}_{\rm odd}}(\bi{I}(X',Y'))}
{\Theta_{\bi{\epsilon}_{\rm odd}}
(\bi{I}(X,Y'))\Theta_{\bi{\epsilon}_{\rm odd}}(\bi{I}(X',Y))}
\end{equation}
is well defined quasi-periodical function on $\Gamma^{\otimes 4}$.
$\bi{\epsilon}_{\rm odd}$ is a non-singular odd theta
characteristic such that $\Theta_{\bi{\epsilon}_{\rm
odd}}(\mathbf{0})=0$. We denote by $\Theta(\bi{v})$ the
theta-function with zero characteristic.

There is a famous identity on $\Gamma^{\otimes 4}\otimes
\rm{Jac}(\Gamma)$,  so called bi-linear Fay's identity, which can
be written in the form:
\begin{equation}\label{Fay}
\begin{array}{l}
\ds \Theta(\bi{v})\;\Theta(\bi{v}+\bi{I}(B+D,A+C))=\\
\quad=\ \ds\frac{E(A,B)E(D,C)}{E(A,C)E(D,B)}\; \Theta(\bi{v} +
\bi{I}(D,A))\;\Theta( \bi{v} + \bi{I}(B,C))+\\ [3mm] \qquad+\
\ds\frac{E(A,D)E(C,B)}{E(A,C)E(D,B)}\; \Theta(\bi{v} +
\bi{I}(B,A))\;\Theta( \bi{v} +\bi{I}(D,C))\ ,
\end{array}
\end{equation}
where one should understand $\bi{I}(B+D,A+C)=\bi{I}(B,A)
+\bi{I}(D,C)=\bi{I}(D,A)+\bi{I}(B,C)$.

Let $\bi{v}\in\mbox{Jac}(\Gamma)$, $X_{n_1}$, $X'_{n_1}$,
$Y_{n_2}$, $Y'_{n_2}$, $Z_{n_3}$, $Z'_{n_3}$
($n_\alpha\in\mathbb{Z}$) and $P,Q$ be arbitrary distinct points on
the algebraic curve $\Gamma$. We have
\begin{prop}\label{prop5}
Any solution of local equations of motion
(\ref{ttt}) is a particular case of the following:
\begin{equation}\label{tau-xyz}
\ds\begin{array}{l} \ds \tau_{1,\pot}\;=\;\xi_{1:\,n_2,n_3}\,
\Theta( \bi{I}_\pot+\bi{I}(Q,X_{n_1}))\;,\\
\\
\ds \tau_{2,\pot}\;=\;\xi_{2:\,n_1,n_3}\, \Theta(
\bi{I}_\pot+\bi{I}(Q,Y_{n_2}))\;,\\
\\
\ds \tau_{3,\pot}\;=\;\xi_{3:\,n_1,n_2}\, \Theta(
\bi{I}_\pot+\bi{I}(Q,Z_{n_3}))\;,
\end{array}
\end{equation}
where
\begin{equation}\label{theta-arg}
\ds \bi{I}_\pot\;=\;\bi{v}\;+\;
\sum_{m_1=0}^{n_1-1}\,\bi{I}(X_{m_1}',X_{m_1}^{}) +
\sum_{m_2=0}^{n_2-1}\,\bi{I}(Y_{m_2}',Y_{m_2}^{}) +
\sum_{m_3=0}^{n_3-1}\,\bi{I}(Z_{m_3}',Z_{m_3}^{})\;,
\end{equation}
and the parameters $\xi$ and points $X_{n_1}^{},...,Z_{n_3}'$
enter the parameterizations of $\kappa_{\alpha,\pot}$ and
pre-exponents as follows:
\begin{equation}\label{kappa-xyz}
\ds \begin{array}{l} \ds \kappa_{1:\,n_2,n_3} \;=\;
-\;\frac{\xi_{1:\,n_2,n_3}\,\xi_{1:\,n_2+1,n_3+1}}
{\xi_{1:\,n_2+1,n_3}\,\xi_{1:\,n_2,n_3+1}}\;
\frac{E(Y_{n_2}',Z_{n_3}^{})\,E(Y_{n_2}^{},Z_{n_3}')}
{E(Y_{n_2}',Z_{n_3}')\,E(Y_{n_2}^{},Z_{n_3}^{})}\;,\\
\\
\ds  \kappa_{2:\,n_1,n_3} \;=\; -\;
\frac{\xi_{2:\,n_1+1,n_3}\,\xi_{2:\,n_1,n_3+1}}
{\xi_{2:\,n_1,n_3}\,\xi_{2:\,n_1+1,n_3+1}}\;
\frac{E(X_{n_1}^{},Z_{n_3}^{})\, E(X_{n_1}',Z_{n_3}')}
{E(X_{n_1}',Z_{n_3}^{})\,
E(X_{n_1}^{},Z_{n_3}')}\;, \\
\\
\ds  \kappa_{3:\,n_1,n_2} \;=\; -\;
\frac{\xi_{3:\,n_1,n_2}\,\xi_{3:\,n_1+1,n_2+1}}
{\xi_{3:\,n_1+1,n_2}\,\xi_{3:\,n_1,n_2+1}}\;
\frac{E(X_{n_1}',Y_{n_2}^{})\, E(X_{n_1}^{},Y_{n_2}')}
{E(X_{n_1}^{},Y_{n_2}^{})\, E(X_{n_1}',Y_{n_2}')}\;,
\end{array}
\end{equation}
and
\begin{equation}\label{uw-xyz}
\ds\begin{array}{l} \ds
\frac{u_{3:\,n_1,n_2}^{(0)}}{w_{1:\,n_2,n_3}^{(0)}}
\;=\; -\; \frac{\xi_{1:\,n_2,n_3}\,\xi_{3:\,n_1,n_2+1}}
{\xi_{1:\,n_2+1,n_3}\,\xi_{3:\,n_1,n_2}}\;
\frac{E(Y_{n_2}',Z_{n_3}^{})\, E(Y_{n_2}^{},X_{n_1}^{})}
{E(Y_{n_2}',X_{n_1}^{})\, E(Y_{n_2}^{},Z_{n_3}^{})}\;,\\
\\
\ds
\frac{u_{2:\,n_1,n_3}^{(0)}}{u_{1:\,n_2,n_3}^{(0)}}
\;=\; \frac{\xi_{1:\,n_2+1,n_3+1}\,\xi_{2:\,n_1,n_3}}
{\xi_{1:\,n_2+1,n_3}\,\xi_{2:\,n_1,n_3+1}}\;
\frac{E(Y_{n_2}',Z_{n_3}^{})\, E(X_{n_1}^{},Z_{n_3}')}
{E(Y_{n_2}',Z_{n_3}')\, E(X_{n_1}^{},Z_{n_3}^{})}\;,\\
\\
\ds
\frac{w_{3:\,n_1,n_2}^{(0)}}{w_{2:\,n_1,n_3}^{(0)}}
\;=\; \frac{\xi_{2:\,n_1+1,n_3}\,\xi_{3:\,n_1,n_2+1}}
{\xi_{2:\,n_1,n_3}\,\xi_{3:\,n_1+1,n_2+1}}\;
\frac{E(Y_{n_2}',X_{n_1}')\, E(X_{n_1}^{},Z_{n_3}^{})}
{E(Y_{n_2}',X_{n_1}^{})\, E(X_{n_1}',Z_{n_3}^{})}\;.
\end{array}
\end{equation}
\end{prop}
\noindent\emph{Proof}: In order to prove that substitution
(\ref{Legendre}) with $\tau$-functions given by (\ref{tau-xyz})
solves (\ref{ttt}) one should use repeatedly  the Fay identity.
For example, equation (\ref{ttt}) for the choice $\alpha=1$,
$\beta=2$ and $\gamma=3$ is the consequence of two Fay identities
(\ref{Fay}) taken for two sets of divisors $(A,B,C,D)$ and
$(A',B',C',D')$ with identification $A=A'=X_{n_1}$,
$B=B'=Y_{n_2}$, $D=D'=Y_{n_2}'$,  $C=Z'_{n_3}$ and $C'=Z_{n_3}$.
The appearing of the ratios containing parameters
$\xi_{\alpha:n_\beta,n_\gamma}$ are due to gauge invariance of the
linear system \r{eq-in} or \r{eq-out}. Gauge parameters
$\xi_{\alpha:n_\beta,n_\gamma}$, the divisors $X_{n_1}$,
$X'_{n_1}$, etc. as well as period matrix $\Omega$ and the point
on the Jacobian  $\bi{v}$ are free parameters of the general
solution. By imposing the periodic boundary condition (which means
in particular the fixing of the size of the system) they may be
related to the values of the integrals of motion.  $\blacksquare$

One may solve expressions (\ref{uw-xyz}) in order to avoid the
rations of the pre-exponents. To do this we   need to introduce
several extra parameters $A_{n_1}$, $B_{n_2}$, $C_{n_3}$. Then,
one obtains for the ``first'' block:
\begin{equation}\label{block1}
\ds \begin{array}{l}
\ds \uop_{1,\pot}\;=\;
\frac{\xi_{1:n_2+1,n_3}}{\xi_{1:n_2+1,n_3+1}}\,
\frac{E(Y_{n_2}',Z_{n_3}')}{E(Y_{n_2}',Z_{n_3}^{})}\,
\frac{E(C_{n_3}^{},Z_{n_3}^{})}{E(C_{n_3}^{},Z_{n_3}')}\,
\frac{\tau_{2,\pot}}{\tau_{2,\pot+\thr}}\;,\\
\\
\ds \wop_{1,\pot}\;=\;-\;
\frac{\xi_{1:n_2+1,n_3}}{\xi_{1:n_2,n_3}}\,
\frac{E(Z_{n_3}^{},Y_{n_2}^{})}{E(Z_{n_3}^{},Y_{n_2}')}\,
\frac{E(B_{n_2}^{},Y_{n_2}')}{E(B_{n_2}^{},Y_{n_2}^{})}\,
\frac{\tau_{3,\pot+\two}}{\tau_{3,\pot}}\;,\\
\\
\ds \kappa_{1:\,n_2,n_3} \;=\;
-\;\frac{\xi_{1:\,n_2,n_3}\,\xi_{1:\,n_2+1,n_3+1}}
{\xi_{1:\,n_2+1,n_3}\,\xi_{1:\,n_2,n_3+1}}\;
\frac{E(Y_{n_2}',Z_{n_3}^{})\,E(Y_{n_2}^{},Z_{n_3}')}
{E(Y_{n_2}',Z_{n_3}')\,E(Y_{n_2}^{},Z_{n_3}^{})}\;,
\end{array}
\end{equation}
for the ``second'' block:
\begin{equation}\label{block2}
\ds \begin{array}{l}
\ds \uop_{2,\pot}\;=\;
\frac{\xi_{2:n_1,n_3}}{\xi_{2:n_1,n_3+1}}\,
\frac{E(X_{n_1}^{},Z_{n_3}')}{E(X_{n_1}^{},Z_{n_3}^{})}\,
\frac{E(C_{n_3}^{},Z_{n_3}^{})}{E(C_{n_3}^{},Z_{n_3}')}\,
\frac{\tau_{1,\pot}}{\tau_{1,\pot+\thr}}\;,\\
\\
\ds \wop_{2,\pot}\;=\;-\;
\frac{\xi_{2:n_1,n_3}}{\xi_{2:n_1+1,n_3}}\,
\frac{E(Z_{n_3}^{},X_{n_1}')}{E(Z_{n_3}^{},X_{n_1}^{})}\,
\frac{E(A_{n_1}^{},X_{n_1}^{})}{E(A_{n_1}^{},X_{n_1}')}\,
\frac{\tau_{3,\pot}}{\tau_{3,\pot+\one}}\;,\\
\\
\ds  \kappa_{2:\,n_1,n_3} \;=\; -\;
\frac{\xi_{2:\,n_1+1,n_3}\,\xi_{2:\,n_1,n_3+1}}
{\xi_{2:\,n_1,n_3}\,\xi_{2:\,n_1+1,n_3+1}}\;
\frac{E(X_{n_1}^{},Z_{n_3}^{})\, E(X_{n_1}',Z_{n_3}')}
{E(X_{n_1}',Z_{n_3}^{})\,
E(X_{n_1}^{},Z_{n_3}')}\;,
\end{array}
\end{equation}
and the ``third'' block finally:
\begin{equation}\label{block3}
\ds \begin{array}{l}
\ds \uop_{3,\pot}\;=\;
\frac{\xi_{3:n_1,n_2+1}}{\xi_{3:n_1,n_2}}\,
\frac{E(X_{n_1}^{},Y_{n_2}^{})}{E(X_{n_1}^{},Y_{n_2}')}\,
\frac{E(B_{n_2}^{},Y_{n_2}')}{E(B_{n_2}^{},Y_{n_2}^{})}\,
\frac{\tau_{1,\pot+\two}}{\tau_{1,\pot}}\;,\\
\\
\ds \wop_{3,\pot}\;=\;-\;
\frac{\xi_{3:n_1,n_2+1}}{\xi_{3:n_1+1,n_2+1}}\,
\frac{E(Y_{n_2}',X_{n_1}')}{E(Y_{n_2}',X_{n_1}^{})}\,
\frac{E(A_{n_1}^{},X_{n_1}^{})}{E(A_{n_1}^{},X_{n_1}')}\,
\frac{\tau_{2,\pot}}{\tau_{2,\pot+\one}}\;,\\
\\
\ds  \kappa_{3:\,n_1,n_2} \;=\; -\;
\frac{\xi_{3:\,n_1,n_2}\,\xi_{3:\,n_1+1,n_2+1}}
{\xi_{3:\,n_1+1,n_2}\,\xi_{3:\,n_1,n_2+1}}\;
\frac{E(X_{n_1}',Y_{n_2}^{})\, E(X_{n_1}^{},Y_{n_2}')}
{E(X_{n_1}^{},Y_{n_2}^{})\, E(X_{n_1}',Y_{n_2}')}\;,
\end{array}
\end{equation}
where $\tau_{\alpha,\pot}$ are given by \r{tau-xyz}. In addition,
the discrete Baker-Akhiezer function $\Phi_{\pot}$, obeying the
whole set of (\ref{eq-in}) (and therefore, (\ref{eq-out})) and
normalized by $\Phi_{\mathbf{0}}\;=\;\xi_{\mathbf{0}}$, is given
by
\begin{equation}\label{baf}
\ds \Phi_{\pot}\;=\;\Phi_\pot(P)\;=\;\xi_{\pot}
\Phi_{\pot}^{(0)}(P)\;\frac{\Theta(\bi{v} + \bi{I}(Q,P) +
\bi{I}_\pot)}{\Theta(\bi{v}+\bi{I}(Q,P))}\;,
\end{equation}
where
\begin{equation}
\begin{array}{rcl}
\ds \Phi^{(0)}_{\pot}(P)&=&\ds
 \prod_{m_1=0}^{n_1-1} \frac{E(P,X_{m_1})E(A_{m_1},X'_{m_1})}
{E(P,X'_{m_1})E(A_{m_1},X_{m_1})}\times\\
&&\ds \times  \prod_{m_2=0}^{n_2-1}
\frac{E(P,Y_{m_2})E(B_{m_2},Y'_{m_2})}
{E(P,Y'_{m_2})E(B_{m_2},Y_{m_2})} \prod_{m_3=0}^{n_3-1}
\frac{E(P,Z_{m_3})E(C_{m_3},Z'_{m_3})}{E(P,Z'_{m_3})E(C_{m_3},Z_{m_3})}
.
\end{array}
\end{equation}

\section{Finite lattice with open boundary}

As one could see,  all the considerations until now were
local and did not take into account the size and the boundary of the
cubic lattice. To describe the global characteristic, such as integrals
of motion, we should specify this external data.

The most natural is the case of the finite cubic lattice with open
boundary conditions and the corresponding Cauchy problem. For the
cubic lattice of the size $N_1\times N_2\times N_3$ we fix the
coordinates $n_1,n_2,n_3$ of 3d vector $\pot$ \r{pot} as follows:
\begin{equation}
\ds 0\leq n_\alpha < N_\alpha\;,\;\;\;\alpha=1,2,3\;.
\end{equation}
$\Delta=N_1N_2+N_2N_3+N_3N_1$ edges incoming to the cubic lattice
correspond to the $2\Delta$ initial data
\begin{equation}\label{in}
\begin{array}{l}
\ds
\uop_{1:n_2,n_3}\equiv\uop_{1,0\one+n_2\two+n_3\thr}\;,\;\;
\wop_{1:n_2,n_3}\equiv\wop_{1,0\one+n_2\two+n_3\thr}\;,\\
\ds
\uop_{2:n_1,n_3}\equiv\uop_{2,n_1\one+0\two+n_3\thr}\;,\;\;
\wop_{2:n_1,n_3}\equiv\wop_{2,n_1\one+0\two+n_3\thr}\;,\\
\ds
\uop_{3:n_1,n_2}\equiv\uop_{3,n_1\one+n_2\two+0\thr}\;,\;\;
\wop_{3:n_1,n_2}\equiv\wop_{3,n_1\one+n_2\two+0\thr}\;.
\end{array}
\end{equation}
Equations of motion \r{eq-of-mo1}--\r{eq-of-mo3}, being applied
recursively, defines the transformation from the initial data
\r{in} to the $2\Delta$ final data
\begin{equation}\label{out}
\begin{array}{l}
\ds
\uop'_{1:n_2,n_3}\equiv\uop_{1,N_1\one+n_2\two+n_3\thr}\;,\;\;
\wop'_{1:n_2,n_3}\equiv\wop_{1,N_1\one+n_2\two+n_3\thr}\;,\\
\ds
\uop'_{2:n_1,n_3}\equiv\uop_{2,n_1\one+N_2\two+n_3\thr}\;,\;\;
\wop'_{2:n_1,n_3}\equiv\wop_{2,n_1\one+N_2\two+n_3\thr}\;,\\
\ds \uop'_{3:n_1,n_2}\equiv\uop_{3,n_1\one+n_2\two+N_3\thr}\;,\;\;
\wop'_{3:n_1,n_2}\equiv\wop_{3,n_1\one+n_2\two+N_3\thr}\;.
\end{array}
\end{equation}
Suppose that the initial data \r{in} are generic. The natural
question arises. How one can invert the parameterizations
described by the formulas \r{block1}--\r{block3} in order to
restore the algebraic geometry  data in terms of \r{in} and the
parameters $\kappa_{\alpha:n_\beta,n_\gamma}$ (the total number of
parameters is equal to $3\Delta$).

In general the solution of this problem is not unique. Evidently,
the same initial data may be parameterized by the infinite number
of the different sets of algebraic geometry data with sufficiently high
genus. But there exists one preferred compact Riemann surface,
which is minimal in the set of all possible curves parameterizing
the dynamics (\ref{eq-of-mo1}-\ref{eq-of-mo3}).

In the Cauchy problem with (\ref{in}) and (\ref{out}) this curve
appears as follows. Let the initial data are parameterized in the
terms of some $\Gamma$. One may write
$$
\ds \uop_{\alpha:n_\beta,n_\gamma}\;=\;
\uop_{\alpha:n_\beta,n_\gamma}(\bi{v})\;,\;\;\;
\wop_{\alpha:n_\beta,n_\gamma}\;=\;
\wop_{\alpha:n_\beta,n_\gamma}(\bi{v})\;.
$$
Then due to (\ref{tau-xyz},\ref{theta-arg}) the solution of the
Cauchy problem is
$$
\ds \uop_{\alpha:n_\beta,n_\gamma}'\;=\;
\uop_{\alpha:n_\beta,n_\gamma}(\bi{v}+\bi{T}_\alpha)\;,\;\;\;
\wop_{\alpha:n_\beta,n_\gamma}\;=\;
\wop_{\alpha:n_\beta,n_\gamma}(\bi{v}+\bi{T}_\alpha)\;,
$$
where
\begin{equation}\label{shift}
\ds \bi{T}_1\;=\;\sum_{n_1=0}^{N_1}\bi{I}(X_{n_1}',X_{n_1}^{})\;,\;\;
\bi{T}_2\;=\;\sum_{n_2=0}^{N_2}\bi{I}(Y_{n_2}',Y_{n_2}^{})\;,\;\;
\bi{T}_3\;=\;\sum_{n_3=0}^{N_3}\bi{I}(Z_{n_3}',Z_{n_3}^{})\;.
\end{equation}
The transformation from the initial data to the final ones is the
\emph{evolution} if
\begin{equation}\label{evol}
\ds \bi{T}_1\;=\;\bi{T}_2\;=\;\bi{T}_3\;=\;\bi{T}\quad
\mbox{mod}\quad\mathbb{Z}^g+\Omega\mathbb{Z}^g.
\end{equation}
In what follows, all the relations of the type \r{evol}
will be understood modulo $\mathbb{Z}^g+\Omega\mathbb{Z}^g$.

The evolution conditions (\ref{evol}) mean that due to the Abel
theorem there exist two meromorphic functions $\lsp(P)$ and
$\msp(P)$, $P\in\Gamma$, with the divisors
\begin{equation}\label{gen-div}
\begin{array}{rcl}
(\lsp)&=&\ds\sum_{n_1\in\mathbb{Z}_{N_1}}X_{n_1}-
\sum_{n_1\in\mathbb{Z}_{N_1}}X'_{n_1}-
\sum_{n_3\in\mathbb{Z}_{N_3}}Z_{n_3}
+\sum_{n_3\in\mathbb{Z}_{N_3}}Z'_{n_3}\;,\\
[3mm] (\msp)&=&\ds\sum_{n_2\in\mathbb{Z}_{N_2}}Y_{n_2}-
\sum_{n_2\in\mathbb{Z}_{N_2}}Y'_{n_2}
-\sum_{n_3\in\mathbb{Z}_{N_3}}Z_{n_3}+
\sum_{n_3\in\mathbb{Z}_{N_3}}Z'_{n_3}\;.\\
\end{array}
\end{equation}
This means that $\Gamma$ is the compact Riemann surface given by a
polynomial equation $J_\triangle(\lambda,\mu)=0$ (see e.g. Theorem
10-23 in \cite{Springer}). Moreover, (\ref{gen-div}) fix the
generic structure of $J_\triangle(\lambda,\mu)$ (later we will
give it, see (\ref{jtriangle})). The key observation is that due
to the evolution conditions (\ref{evol})
$J_\triangle(\lambda,\mu)$ as a functional of the initial data
should produce the set of invariants of the evolution, so the
curve is the spectral one.

$J_\triangle$ may be derived  with the help of the linear system
of the type \r{eq-in} written for the whole auxiliary plane. Let
us fix the position of the auxiliary planes corresponding to the
initial and final data. The auxiliary plane for the initial data
crosses all the incoming edges of the cubic lattice while for the
final data it intersects all the outgoing edges. These auxiliary
planes plays the role of the two-dimensional  space-like surfaces
and discrete evolution corresponds to the translation of the
auxiliary plane into direction perpendicular to this space-like
surfaces.

\begin{figure}
\setlength{\unitlength}{0.0004in}
\begin{center}
{\renewcommand{\dashlinestretch}{30}
\begin{picture}(10182,9710)(0,-10)
\path(270,1133)(10170,1133)
\whiten\path(10050.000,1103.000)(10170.000,1133.000)
(10050.000,1163.000)(10050.000,1103.000)
\path(270,2483)(10170,2483)
\whiten\path(10050.000,2453.000)(10170.000,2483.000)
(10050.000,2513.000)(10050.000,2453.000) \path(495,233)(6795,8558)
\whiten\path(6746.509,8444.208)(6795.000,8558.000)
(6698.665,8480.415)(6746.509,8444.208) \path(9315,233)(5445,9683)
\whiten\path(5518.239,9583.320)(5445.000,9683.000)
(5462.715,9560.582)(5518.239,9583.320) \path(1857,179)(7932,8054)
\whiten\path(7882.457,7940.662)(7932.000,8054.000)
(7834.950,7977.310)(7882.457,7940.662) \path(8019,12)(4419,9057)
\whiten\path(4491.249,8956.600)(4419.000,9057.000)
(4435.502,8934.413)(4491.249,8956.600)
%
\put(6075,8783){\makebox(0,0)[lb]{$\Phi_{1:22}$}}
\put(5200,7520){\makebox(0,0)[lb]{$\Phi_{1:21}$}}
\put(6660,7523){\makebox(0,0)[lb]{$\Phi_{1:12}$}}
\put(5760,6488){\makebox(0,0)[lb]{$\Phi_{1:11}$}}
\put(4455,6578){\makebox(0,0)[lb]{$\Phi_{1:20}$}}
\put(7245,6578){\makebox(0,0)[lb]{$\Phi_{1:02}$}}
\put(4860,5498){\makebox(0,0)[lb]{$\Phi_{1:10}$}}
\put(6300,5453){\makebox(0,0)[lb]{$\Phi_{1:01}$}}
\put(5400,4350){\makebox(0,0)[lb]{$\Phi_{1:00}$}}
\put(4230,2753){\makebox(0,0)[lb]{$\Phi_{3:00}$}}
\put(2925,2798){\makebox(0,0)[lb]{$\Phi_{3:01}$}}
\put(1530,2843){\makebox(0,0)[lb]{$\Phi_{3:02}$}}
\put(765,1808){\makebox(0,0)[lb]{$\Phi_{3:12}$}}
\put(2160,1763){\makebox(0,0)[lb]{$\Phi_{3:11}$}}
\put(3645,1763){\makebox(0,0)[lb]{$\Phi_{3:10}$}}
\put(   0,683){\makebox(0,0)[lb]{$\Phi_{3:22}$}}
\put(1350,683){\makebox(0,0)[lb]{$\Phi_{3:21}$}}
\put(2835,638){\makebox(0,0)[lb]{$\Phi_{3:20}$}}
\put(5950,2753){\makebox(0,0)[lb]{$\Phi_{2:00}$}}
\put(6400,1718){\makebox(0,0)[lb]{$\Phi_{2:10}$}}
\put(6900,548){\makebox(0,0)[lb]{$\Phi_{2:20}$}}
\put(7380,2753){\makebox(0,0)[lb]{$\Phi_{2:01}$}}
\put(7740,1673){\makebox(0,0)[lb]{$\Phi_{2:11}$}}
\put(8190,503){\makebox(0,0)[lb]{$\Phi_{2:21}$}}
\put(8685,2753){\makebox(0,0)[lb]{$\Phi_{2:02}$}}
\put(9045,1628){\makebox(0,0)[lb]{$\Phi_{2:12}$}}
\put(9585,503){\makebox(0,0)[lb]{$\Phi_{2:22}$}}
\put(-1000,1000){\makebox(0,0)[lb]{$\sk{n_1=1}$}}
\put(   0,-200){\makebox(0,0)[lb]{$\sk{n_2=1}$}}
\put(1500,-200){\makebox(0,0)[lb]{$\sk{n_2=0}$}}
\put(-1000,2300){\makebox(0,0)[lb]{$\sk{n_1=0}$}}
\put(7600,-200){\makebox(0,0)[lb]{$\sk{n_3=0}$}}
\put(9000,-200){\makebox(0,0)[lb]{$\sk{n_3=1}$}}
\end{picture}
}
\caption{ \label{ex-of-enum}\ftn{Example of enumeration of the linear
variables.}}
\end{center}
\end{figure}
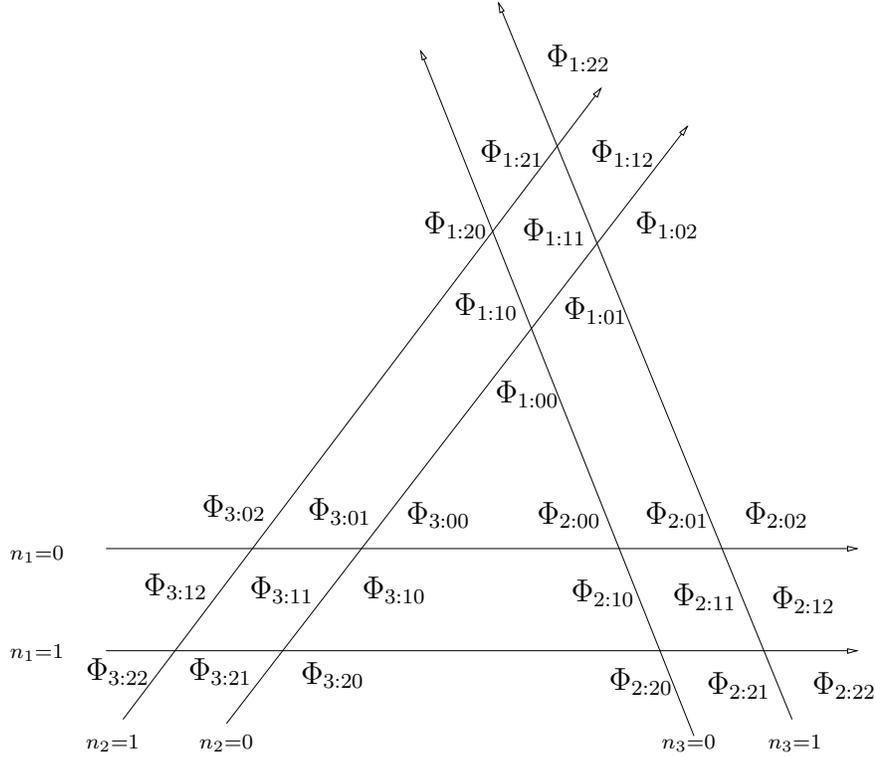

All the objects, namely the dynamical and auxiliary linear
variables on the space-like surface, can be numbered by the
two-dimensional discrete index. Let us choose the numbering for
the linear variables in the form similar to numbering of the
initial \r{in} or final \r{out} data:
\begin{equation}\label{linear-var}
\begin{array}{l}
\ds \Phi_{0\one+n_2\two+n_3\thr}\;=\;\Phi_{1:n_2,n_3}\,,\;\;\;
\Phi_{n_1\one+0\two+n_3\thr}\;=\;\Phi_{2:n_1,n_3}\,,\\
\\
\ds \Phi_{n_1\one+n_2\two+0\thr}\;=\; \Phi_{3:n_1,n_2}\,,\;\;\;
0\leq n_i \leq N_i\,,\;\;\; i=1,2,3\;.
\end{array}
\end{equation}
An example
of such enumeration on the auxiliary plane
in the simple case $N_1=N_2=N_3=2$ is shown in
Fig.~\ref{ex-of-enum}.

Let $\mathcal{L}(\lambda,\mu)$ be the matrix of the linear system
\begin{equation}\label{lin-ge-cl}
\left\{\begin{array}{rcl}
0&=&\Phi_{1:n_2+1,n_3}-\Phi_{1:n_2+1,n_3+1}u_{1:n_2,n_3}+
\Phi_{1:n_2,n_3}w_{1:n_2,n_3}+ \\
&& + \ \Phi_{n_2,n_3+1}\kappa_{1:n_2,n_3}u_{1:n_2,n_3}
w_{1:n_2,n_3\;,}\\
0&=&\Phi_{2:n_1,n_3}-\Phi_{2:n_1,n_3+1}u_{2:n_1,n_3}+
\Phi_{2:n_1+1,n_3}w_{2:n_1,n_3}+ \\
&& + \ \Phi_{2:n_1+1,n_3+1}
\kappa_{2:n_1,n_3}u_{2:n_1,n_3}
w_{2:n_1,n_3}\;,\\
0&=&\Phi_{3:n_1,n_2+1}-\Phi_{3:n_1,n_2}u_{3:n_1,n_2}+
\Phi_{3:n_1+1,n_2+1}w_{3:n_1,n_2}+ \\
&& + \ \Phi_{3:n_1+1,n_2} \kappa_{3:n_1,n_2}u_{3:n_1,n_2}
w_{3:n_1,n_2}
\end{array}\right.
\end{equation}
written in the matrix form $0=\Phi \cdot \mathcal{L}(\lsp,\msp)$,
 where the linear variables satisfy the identification
conditions
\begin{equation}\label{phi-ident}
\ds \Phi_{1:0,n_3}=\Phi_{2:0,n_3}\;,\;\;\;
\Phi_{1:n_2,0}=\Phi_{3:0,n_2}\;,\;\;\;
\Phi_{2:n_1,0}=\Phi_{3:n_1,0}\;,
\end{equation}
and the quasi-periodicity conditions
\begin{equation}\label{phi-period}
\ds\frac{\Phi_{3:N_1,n_2}}{x}=\frac{\Phi_{1:n_2,N_3}}{z}\;,\;\;\;
\frac{\Phi_{3:n_1,N_2}}{y}=\frac{\Phi_{2:n_1,N_3}}{z}\;,\;\;\;
\frac{\Phi_{1:N_2,n_3}}{y}=\frac{\Phi_{2:N_1,n_3}}{x}\;
\end{equation}
for the boundary domains on the auxiliary plane. Parameters
\begin{equation}\label{spec-par}
\lsp=\frac{x}{z}\;,\qquad \msp=\frac{y}{z}
\end{equation}
are the complex numbers and we call them the  spectral parameters.
Due to \r{phi-ident} and \r{phi-period} it is clear that the total
number of the independent linear variables in the system
\r{lin-ge-cl} is $\Delta$ and $\mathcal{L}(\lsp,\msp)$ is
$\Delta\times\Delta$ square matrix.

\begin{figure}
\setlength{\unitlength}{0.0004in}
\begin{center}
{\renewcommand{\dashlinestretch}{30}
\begin{picture}(4756,2883)(0,-10)
\path(405,381)(4680,381)
\blacken\path(4560.000,351.000)(4680.000,381.000)
(4560.000,411.000)(4560.000,351.000)
\path(630,156)(630,2856)
\blacken\path(660.000,2736.000)
(630.000,2856.000)(600.000,2736.000)(660.000,2736.000)
\dashline{60.000}(4230,381)(4230,1281)
\dashline{60.000}(630,2406)(1530,2406)
\thicklines
\path(630,381)(630,1506)(1530,2406)
    (4230,2406)(4230,1281)(3330,381)(630,381)
\put(-400,2271){\makebox(0,0)[lb]{$\sk{N_2+N_3}$}}
\put(200,1416){\makebox(0,0)[lb]{$\sk{N_2}$}}
\put(3000,21){\makebox(0,0)[lb]{$\sk{N_1}$}}
\put(4000,21){\makebox(0,0)[lb]{$\sk{N_1+N_3}$}}
\put(4680,561){\makebox(0,0)[lb]{$b$}}
\put(810,2766){\makebox(0,0)[lb]{$a$}}
\end{picture}
}
\caption{ \label{Newton-pol}\ftn{Newton's polygon.}}
\end{center}
\end{figure}
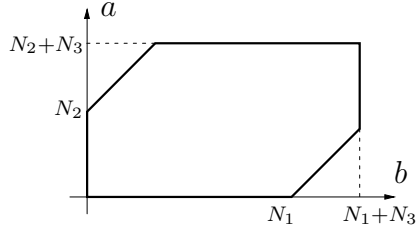

Define
\begin{equation}\label{determin}
J_\triangle(\lsp,\msp)=\det \mathcal{L}(\lsp,\msp)\,
\left(\prod_{n_2,n_3}\,\uop_{1:n_2,n_3}\right)^{-1}\;.
\end{equation}
\r{determin} is a normalized  Laurent polynomial ($J_{0,0}=1$)
of the spectral parameters $\lsp$ and $\msp$,
\begin{equation}\label{jtriangle}
\ds J_\triangle=\sum_{a,b\in\Pi}\lsp^a\msp^{-b} J_{a,b}\;,\;\;\;
\Pi\;\;:\;\;\left\{\begin{array}{l} 0\leq a\leq N_2+N_3\\
0\leq b \leq N_1+N_3\\
-N_1\leq a-b\leq N_2
\end{array}\right.\;.
\end{equation}
 Domain $\Pi$ (Newton's polygon of
$J_\triangle$) is shown in Fig.~\ref{Newton-pol}. One may prove
(see \cite{Sergeev}) that the coefficients of this Laurent
polynomial are invariants of the evolution, i.e they are the same
for the initial \r{in} and final \r{out} data and can be served as
a source of the algebraic geometry data for the solution of these
equations with open boundary conditions.

The requirement that the linear system \r{lin-ge-cl} has
non-trivial solution is equivalent to the fact that the spectral
parameters belongs to the algebraic curve:
\begin{equation}\label{cl-Big-curve}
P=(\lsp,\msp)\in\Gamma_{\triangle}\;\;
\Leftrightarrow\;J_\triangle(\lsp,\msp)=0\;.
\end{equation}
Assuming that all the incoming data together with parameters
$\kappa_{\alpha:n_\beta,n_\gamma}$ are in general position, one
may calculate the genus of the curve (\ref{cl-Big-curve}) using
Newton's polygon. For the cubic lattice of the size $N_1\times
N_2\times N_3$ such that $N_1\geq N_2\geq N_3$ Newton's polygon
associated with this curve \r{cl-Big-curve} is shown in
Fig.~\ref{Newton-pol}.
The genus of the curve $\Gamma_\triangle$ is equal to the number
of internal points of this polygon
\begin{equation}\label{genus0}
g_\triangle=(N_1N_2+N_1N_3+N_2N_3)-(N_1+N_2+N_3)+1\;.
\end{equation}
$g$ coefficients $J_{a,b}$ in the Laurent polynomial
$J_\triangle(\lsp,\msp)$ corresponding to the internal points of
the Newton's polygon are related to the moduli of the algebraic
curve $\Gamma_\triangle$ and so give the period matrix
$\Omega_\triangle$. Coefficients $J_{a,b}$ corresponding to the
perimeter of the polygon are related to the divisors of the
meromorphic on $\Gamma_\triangle$ functions $\lsp,\msp$ and so
give the set of $X_{n_1}^{},...,Z_{n_3}'$. From (\ref{gen-div}) and
 (\ref{baf}) we may get explicitly
\begin{equation}
\begin{array}{l}
\ds\lambda\;=\;\prod_{n_1=0}^{N_1-1}
\frac{E(P,X_{n_1})E(A_{n_1},X'_{n_1})}
{E(P,X'_{n_1})E(A_{n_1},X_{n_1})} \prod_{n_2=0}^{N_2-1}
\frac{E(P,Y_{n_2}')E(B_{n_2},Y_{n_2})}
{E(P,Y_{n_2})E(B_{n_2},Y_{n_2}')}\;,\\
\\
\ds \mu\;=\;\prod_{n_2=0}^{N_2-1}
\frac{E(P,Y_{n_2})E(B_{n_2},Y_{n_2}')}
{E(P,Y_{n_2}')E(B_{n_2},Y_{n_2})} \prod_{n_3=0}^{N_3-1}
\frac{E(P,Z_{n_3}')E(C_{n_3},Z_{n_3})}
{E(P,Z_{n_3})E(C_{n_3},Z_{n_3}')}\;.
\end{array}
\end{equation}

Previously in \cite{SNMP} a bit different approach was used. Due
to the Theorem 2 of \cite{SNMP}, the solution $\Phi(\lambda,\mu)$
of the linear problem $\Phi\cdot\mathcal{L}(\lambda,\mu)=0$ as the
vector of meromorphic functions on the curve $\Gamma_\triangle$ is
given by (\ref{baf}) on $\mathrm{Jac}(\Gamma_\triangle)$. With the
help of (\ref{baf}) the parameterizations
(\ref{tau-xyz}-\ref{uw-xyz}) may be restored uniquely.

We conclude the discussion on the Cauchy problem
by the calculation of the degrees
of freedom of the algebraic geometry parameterizations. With
$\Delta=N_1N_2+N_2M_3+N_3N_1$, $\Delta'=N_1+N_2+N_3$ and
$g=\Delta-\Delta'+1$, we  have $g$ moduli of $\Gamma_\triangle$,
$g$ complex numbers $\bi{v}\in\mathrm{Jac}(\Gamma_\triangle)$, $g$
independent cross-ratios of $\xi_{\pot}$, $2\Delta'-2$ independent
divisors $X_{n_1}^{},...,Z_{n_3}'$ and $\Delta'$ arbitrary
divisors $A_{n_1},B_{n_2},C_{n_3}$ \emph{minus} one degree of
freedom corresponding to the arbitrariness of $Q$: totally
$3\Delta$ parameters. It proves the completeness of the
parameterizations in the terms of $\Gamma_\triangle$.

\section{B\"acklund transformation}
\label{integrability}

Curve $\Gamma_\triangle$ corresponds to the generic position of
the initial data and open boundary conditions. For the application
to the integrable spin models the periodical boundary conditions
$\bi{T}=0$, i.e.
\begin{equation}\label{xyz-period}
\ds \sum_{n_1\in\mathbb{Z}_{N_1}}\bi{I}(X_{n_1}',X_{n_1}^{})\;=\;
\sum_{n_2\in\mathbb{Z}_{N_2}}\bi{I}(Y_{n_2}',Y_{n_2}^{})\;=\;
\sum_{n_3\in\mathbb{Z}_{N_3}}
\bi{I}(Z_{n_3}',Z_{n_3}^{})\;=\;0
\end{equation}
are important. Periodical boundary conditions reduces the
spectral curve $\Gamma_\triangle$. Namely, due to \r{xyz-period},
the parameters $x,y,z$ in
(\ref{phi-period}) become the meromorphic functions with divisors
$(x)=\sum_{n_1}X^{}_{n_1}-\sum_{n_1}X'_{n_1}$ and \r{yz-divisors}.
Since now $x,y,z$ are meromorphic functions (not only their ratios)
the curve can be defined by an algebraic equations for any pair
from $x,y,z$:
\begin{equation}\label{reduction}
J_1(y,z)\;=\;J_2(x,z)\;=\;J_3(x,y)\;=\;0\;.
\end{equation}
The Laurent polynomial  $J_\triangle(x/z,y/z)$ is not a generic one
and $J_\triangle=0$ is a consequence of the algebraic relations \r{reduction}.

In this section we describe explicitly the reduced spectral curve
and the meromorphic functions, which uniformize it for another
choice of the auxiliary plane. The interpretation of the problem
now differs from the Cauchy one for the open cubic lattice.
The discrete evolution can be identified  now
with a sequence of the B\"acklund transforms for the
square auxiliary plane.

First we define the new position of the auxiliary plane. Let it
crosses the ingoing edges for the vertices of the 3-dimensional
lattice with coordinates $n_2\two+n_3\thr$. With this choice the
role of the variables $\uop_\alpha$ and $\wop_\alpha$,
$\alpha=1,2,3$ becomes different: $\uop_{2,\pot}$,
$\wop_{2,\pot}$, $\uop_{3,\pot}$, $\wop_{3,\pot}$ are auxiliary,
while $\uop_{1,\pot},\wop_{1,\pot}$ are dynamical. The evolution
will describe the change of these dynamical variables with respect
to the "time" $n_1$ with periodic boundary conditions implied in
the second and third directions:
\begin{equation}\label{per23}
\uop_{1:n_2+N_2,n_3}=\uop_{1:n_2,n_3+N_3}=\uop_{1:n_2,n_3}\;,\quad
\wop_{1:n_2+N_2,n_3}=\wop_{1:n_2,n_3+N_3}=\wop_{1:n_2,n_3}\;.
\end{equation}
Of course, the auxiliary for this evolution variables are also
different from layer to layer according to the discrete equations
\r{eq-of-mo2}--\r{eq-of-mo3}. We will keep the open boundary
conditions in the first direction.

Consider one layer of the cubic lattice corresponding to $n_1=0$.
Equations of motion \r{eq-of-mo1} and the periodical boundary
condition \r{per23} yields an implicit transformation
\begin{equation}\label{Baeck}
\ds \uop_{1:n_2,n_3},\wop_{1:n_2,n_3}\mapsto
\uop'_{1:n_2,n_3},\wop'_{1:n_2,n_3}.
\end{equation}
According to (\ref{tau-xyz}-\ref{uw-xyz}), one may define
\begin{equation}
\ds u_{1:n_2,n_3}=u_{1:n_2,n_3}(\bi{v})\;,\;\;\;
w_{1:n_2,n_3}=w_{1:n_2,n_3}(\bi{v})\;,
\end{equation}
where $\bi{v}\in\mbox{Jac}(\Gamma)$ for some
algebraic curve $\Gamma$. Then the transformation
\r{Baeck} can be written in the form
\begin{equation}
\ds u'_{1:n_2,n_3}=u_{1:n_2,n_3}(\bi{v}+\bi{I}(X'_0,X_0))\;,\;\;\;
w'_{1:n_2,n_3}=w_{1:n_2,n_3}(\bi{v}+\bi{I}(X'_0,X_0))\;.
\end{equation}
The periodicity conditions (\ref{per23}) in the parameterizations
(\ref{tau-xyz}-\ref{uw-xyz}) means
$\sum_{n_2\in\mathbb{Z}_{N_2}}\bi{I}(Y_{n_2}'Y_{n_2}^{})
=\sum_{n_3\in\mathbb{Z}_{N_3}}\bi{I}(Z_{n_3}',Z_{n_3}^{})=0$,
(see (\ref{xyz-period})). Repeating the consideration from page
\pageref{gen-div}, we may conclude that there exist two
meromorphic functions $y$ and $z$ on $\Gamma$ such that
\begin{equation}\label{yz-divisors}
\ds (y)\;=\;\sum_{n_2}Y_{n_2}^{}-Y_{n_2}'\;,\;\;\;
(z)\;=\;\sum_{n_3} Z_{n_3}^{}-Z_{n_3}'\;,
\end{equation}
and a compact Riemann surface $\Gamma$ may be defined by a
polynomial equation $J_\Box(y,z)=0$. Moreover, the structure of
$(y)$ and $(z)$ fixes the algebraic form of $\Gamma$ (see (\ref{gener})
later).

We call transformation (\ref{Baeck}) the B\"acklund transformation
for the two-dimensional square lattice, since \r{Baeck} is a
canonical transformation and it conserve the integrals of motion.
In order to specify the integrals of motion we again consider the
linear system for the chosen  auxiliary plane. This linear system
is a system of $N_2 N_3$ equations
\begin{equation}\label{system}
j_{n_2,n_3}=0,\quad \Phi_{-1,n_3} = y^{-1}\Phi_{N_2-1,n_3}\;,\quad
\Phi_{n_2,N_3} = z\Phi_{n_2,0}\;,\quad \Phi_{-1,N_3} =
y^{-1}z\Phi_{N_2-1,0}\;,\quad
\end{equation}
where $0\leq n_2<N_2$, $0\leq n_3<N_3$ and linear form
$j_{n_2,n_3}$ is defined by the first line in \r{eq-in} with
slightly shifted enumeration of the auxiliary linear variables
$\Phi_{n_2,n_3}$
\begin{equation}\label{lin-Phi}
\begin{array}{rcl}
\ds j_{n_2,n_3}&=&\Phi_{n_2,n_3} - \Phi_{n_2,n_3+1}
u_{1:\,n_2,n_3} + \Phi_{n_2-1,n_3} w_{1:\,n_2,n_3}\ +\\
&&+\ \Phi_{n_2-1,n_3+1} \kappa_{1:\,n_2,n_3} u_{1:\,n_2,n_3}
w_{1:\,n_2,n_3}\;.
\end{array}
\end{equation}
Equations in \r{system} which includes the spectral parameters $y$
and $z$ describe quasi-periodical boundary conditions for the
linear variables $\Phi_{n_2,n_3}$.

Let $L(y,z)$ be the complete matrix of the coefficients of the
system \r{system}, $j=\Phi\cdot L(y,z)$. Define a Laurent polynomial
$J_\Box(y,z)$
\begin{equation}\label{gener}
J_\Box(y,z)=\det\;L(y,z)=\sum_{a=0}^{N_3}\sum_{b=0}^{N_2}
J_{a,b}\;y^{-a}\;z^b\;,\;\;\;J_{0,0}\;=\;1\;.
\end{equation}
$J_{a,b}$ in this case are the invariants of the B\"acklund
transform (\ref{Baeck}), see \cite{Sergeev}. In order the system
$\Phi(P)\cdot L(y,z)=0$ has a nontrivial solution, the spectral parameters
$y$ and $z$ should belong to the spectral curve
\begin{equation}\label{curve-sm}
P=(y,z)\in\Gamma_\Box\;\;\Leftrightarrow\;\; J_\Box(y,z)=0\;.
\end{equation}
Assuming again that the dynamical variables $\uop_{1:n_2,n_3}$,
$\wop_{1:n_2,n_3}$ and parameters $\kappa_{1:n_2,n_3}$ are
generic, the genus of the spectral curve \r{curve-sm} is equal to
$g_\Box=(N_2-1)(N_3-1)$. Recall, the form (\ref{gener}) of the
polynomial $J_\Box$ follows uniquely from the conditions
$(y)=\sum Y_{n_2}^{}-Y_{n_2}'$, $(z)=\sum Z_{n_3}^{}-Z_{n_3}'$.

In the backward direction, Theorem 2 from \cite{SNMP} says that
the non-normalized solution
of $\Phi(P)\cdot L(y,z)=0$ as the meromorphic function
of $\Gamma_\Box$ is given by
\begin{equation}\label{sol-23}
\begin{array}{rcl}
\ds\Phi_{n_2,n_3}(P)&=&\ds
\prod_{m_2=0}^{n_2-1}
\frac{E(P,Y_{m_2})}{E(P,Y'_{m_2})}
\frac{E(B_{m_2},Y'_{m_2})}{E(B_{m_2},Y_{m_2})}\ \
\prod_{m_3=0}^{n_3-1}
\frac{E(P,Z_{m_3})}{E(P,Z'_{m_3})}
\frac{E(C_{m_3},Z'_{m_3})}{E(C_{m_3},Z_{m_3})}\times\\
&&\ds\times
\Theta\left(
\ds\bi{v}+
\bi{I}(Q,P) +
\sum_{m_2=0}^{n_2-1}\,\bi{I}(Y_{m_2}',Y_{m_2}^{}) +
\sum_{m_3=0}^{n_3-1}\,\bi{I}(Z_{m_3}',Z_{m_3}^{})
\right),
\end{array}
\end{equation}
where theta-functions are constructed by means of the period
matrix $\Omega_\Box$ of the algebraic curve $\Gamma_\Box$. Using
formula \r{sol-23} we may find the corresponding expressions for
the dynamical variables $u_{1:n_2,n_3}$ and $w_{1:n_2,n_3}$ to
observe that they coincide with those given by the
Proposition~\ref{prop5} and arbitrary curve $\Gamma$ being identified
with the spectral curve $\Gamma_\Box$. Explicit form of the spectral
parameters uniformizing the spectral curve $\Gamma_\Box$  are
\begin{equation}\label{func-YZ}
\begin{array}{rcl}
\ds z(P)&=&\ds\prod_{n_3\in\mathbb{Z}_{N_3}}
\frac{E(P,Z_{n_3})}{E(P,Z_{n_3}')}\;
 \frac{E(C_{n_3},Z_{n_3}')}{E(C_{n_3},Z_{n_3})}\;,\\
 &&\\
y(P)&=&\ds \prod_{n_2\in\mathbb{Z}_{N_2}}
\frac{E(P,Y_{n_2})}{E(P,Y_{n_2}')}\;
\frac{E(B_{n_2},Y_{n_2}')}{E(B_{n_2},Y_{n_2})}\;.
\end{array}
\end{equation}

In case when the algebraic curve used for the construction of the
general solution becomes a rational one, the one step of this
evolution $n_1\mapsto n_1+1$ with open b.c. in the $1$-st
direction can be identified with creation of additional soliton.
This will be explicitly demonstrated in the next section.

In this Section we have established the origin of the
curve $\Gamma_\Box$ appearing when the periodical b.c. are imposed
in $2$-nd and $3$-rd directions. Evidently, $J_\Box(y,z)$ should
be identified with $J_1(y,z)$ in (\ref{reduction}). In the same
way the periodical b.c. may be imposed in any other pair of
directions, and the corresponding $J_2$ and $J_3$ also are the
spectral determinants. If the periodical b.c. are imposed in all
three directions, then any of three relations $J_1=0$, $J_2=0$ and
$J_3=0$ should define the same curve, and since the principle of
generic data in this case has been lost, the genus of the curve
may be established as
\begin{equation}
\ds g \;\leq\; \textrm{min of }\;(N_1-1)(N_2-1)\;,\;\;
(N_2-1)(N_3-1)\;,\;\; (N_3-1)(N_1-1)\;.
\end{equation}

\section{Rational limit}

Let us again forget on the boundary conditions and consider a
rational limit of the algebraic geometry solutions to the discrete
equation of motion \r{ttt}. This rational limit for the
$\Theta$-function (\ref{Theta}) on a Jacobian of an algebraic
curve corresponds to (see \cite{RTC})
\begin{equation}
\ds \EXP^{i\pi\Omega_{n,n}+2 i \pi v_n}\;=\;-f_n^{}\;,
\end{equation}
and
\begin{equation}
\ds \EXP^{i\pi\Omega_{n,n}}\;\mapsto\;0\;,\;\;\;
\EXP^{i\pi\Omega_{k,n}}\;\mapsto\; \frac{(q_k-q_n)\,(p_k-p_n)}
{(q_k-p_n)\,(p_k-q_n)}\;\stackrel{def}{=}\;d_{k,n}\;.
\end{equation}
Thus the set of $d_{k,n}$ is the reminder of the period matrix,
while
the set of $f_n$ is the reminder of $\bi{v}$. The prime forms in
the rational limit are the prime forms on the sphere:
\begin{equation}\label{prime-rat}
\ds \frac{E(A,B)E(C,D)}{E(A,D)E(C,B)} \;=\;
\frac{(A-B)(C-D)}{(A-D)(C-B)}\;.
\end{equation}
For $g=1,2,\ldots$ and the sequence of the parameters
\begin{equation}\label{sequence}
\ds p_0,q_0,f_0;\;\;p_2,q_2,f_2;
\;\;...\;\;p_{g-1},q_{g-1},f_{g-1}\;=\;\{p_k,q_k,f_k\}_{k=0}^{g-1}
\end{equation}
we introduce the rational limit of the $\Theta$-function (see the
appendix of \cite{RTC})
\begin{equation}\label{casor}
\ds H^{(g)}(\{p_k,q_k,f_k\}_{k=0}^{g-1})\;=\;\frac{\ds \det
|q_j^{i}\,-\,f_j^{}\,p_j^{i}|_{i,j=0}^{g-1}}
{\ds\prod_{i>j}\,(q_i^{}\,-\,q_j^{})}\;.
\end{equation}
We set $H^{(0)}\equiv1$ by definition. Note that if all
parameters $f_k$ vanish $H^{(g)}(\{p_k,q_k,0\}_{k=0}^{g-1})=1$
as well. Let the function
$\sigma_k(z)$ be
\begin{equation}\label{S-fun}
\ds \sigma_k(z)\;=\; \frac{p_k\,-\,z}{q_k\,-\,z}\;.
\end{equation}

When $g=0$ the equations \r{ttt} in the rational limit have
the simple solution
$\tau_{\alpha,\pot}=1$ due to identity
\begin{equation}\label{sim-id}
r_\alpha=1+s_\beta+s^{-1}_\gamma\;,
\end{equation}
where
\begin{equation}\label{sol-par}
\begin{array}{rcl}
r_\alpha&=&\ds -\frac{X'_\beta-X'_\gamma}{X_\beta-X_\gamma}\;
\frac{X_\alpha-X_\beta}{X_\alpha-X_\gamma}\;
\frac{X_\alpha-X'_\beta}{X_\alpha-X'_\gamma}\;,\\
&&\\
s_\alpha&=&\ds - \frac{X_\alpha-X_\gamma}{X_\alpha-X_\beta}\;
\frac{X'_\alpha-X_\beta}{X'_\alpha-X_\gamma}
\end{array}
\end{equation}
and $(\alpha,\beta,\gamma)$ is any even permutation of the set
$(1,2,3)$. The notations in \r{sim-id} and \r{sol-par}
are related to those in \r{rs-uwk} and the Proposition~\ref{prop5}
as follows: $r_\alpha=r_{\alpha,\pot}$, $s_\alpha=s_{\alpha,\pot}$;
$X_1=X_{n_1}$, $X'_1=X'_{n_1}$, $X_2=Y_{n_2}$,
$X'_2=Y'_{n_2}$, $X_3=Z_{n_3}$ and finally $X'_3=Z'_{n_3}$.

\begin{prop}\label{prop44}
The soliton solutions of the equation \r{ttt} are given
by the formulas
\begin{equation}\label{tau-xyz-rat}
\ds\begin{array}{rcl}
\ds \tau_{1,\,\pot}&=&\ds
H^{(g)}\left(\left\{\frac{I_{\pot:\,k}}{\sigma_k(X_{n_1})}
\right\}_{k=0}^{g-1}\right)
\;,\\
&&\\
\ds \tau_{2,\,\pot} &=&\ds
H^{(g)}\left(\left\{\frac{I_{\pot:\,k}}
{\sigma_k(Y_{n_2})}\right\}_{k=0}^{g-1}\right)
\;,\\
&&\\
\ds \tau_{3,\,\pot} &=&\ds
H^{(g)}\left(\left\{\frac{I_{\pot:\,k}}
{\sigma_k(Z_{n_3})}\right\}_{k=0}^{g-1}\right) \;,
\end{array}
\end{equation}
where $I_{\pot:\,k}$, $n_1,n_2,n_3\in\mathbb{Z}$
\begin{equation}\label{imbedding-rat}
\ds I_{\pot:\,k} \;=\; f_k\;\left(\prod_{m_1=0}^{n_1-1}\,
\frac{\sigma_k(X_{m_1}')}{\sigma_k(X_{m_1}^{})}\right)
\left(\prod_{m_2=0}^{n_2-1}\,
\frac{\sigma_k(Y_{m_2}')}{\sigma_k(Y_{m_2}^{})}\right)
\left(\prod_{m_3=0}^{n_3-1}\,
\frac{\sigma_k(Z_{m_3}')}{\sigma_k(Z_{m_3}^{})}\right)\;,
\end{equation}
and parameters $r_{\alpha,\pot}$ and $s_{\alpha,\pot}$ are
defined by the formulas \r{sol-par}.
\end{prop}

The proof of this Proposition is based on the rational variant
of the Fay's identity
\begin{equation}\label{fay-identity-rat}
\ds\begin{array}{l} \ds (A-D)(C-B)\,
H^{(g)}(\{f_k\frac{\sigma_k(A)}{\sigma_k(B)}\}_{k=1}^g)\,
H^{(g)}(\{f_k\frac{\sigma_k(C)}{\sigma_k(D)}\}_{k=1}^g)\;+ \\
\\
\ds (A-B)(D-C)\,
H^{(g)}(\{f_k\frac{\sigma_k(A)}{\sigma_k(D)}\}_{k=1}^g)\,
H^{(g)}(\{f_k\frac{\sigma_k(C)}{\sigma_k(B)}\}_{k=1}^g)\;= \\
\\
\ds (A-C)(D-B)\, H^{(g)}(\{f_k\}_{k=1}^g)\,
H^{(g)}(\{f_k\frac{\sigma_k(A)\sigma_k(C)}
{\sigma_k(B)\sigma_k(D)}\}_{k=1}^g)
\end{array}
\end{equation}
described in \cite{RTC}.

So far  the parameters $p_k$, $q_k$ and $f_k$ \r{sequence} are
arbitrary complex parameters and the solution given by the
Proposition~\ref{prop44} is relevant for lattice infinite in all
directions. Now we would like to impose periodic boundary
conditions \r{per23} in directions 2 and 3 and interpret the
evolution along the direction 1 as a sort of B\"acklund
transformation which create   the solitons. To do this we first
note that  boundary conditions \r{per23} are equivalent to
the following algebraic relations for the parameters $p$ and $q$:
\begin{equation}\label{per23-sol}
\prod_{n_2=0}^{N_2-1}\frac{\sigma(Y'_{n_2})}{\sigma(Y_{n_2})}=
\prod_{n_3=0}^{N_3-1}\frac{\sigma(Z'_{n_3})}{\sigma(Z_{n_3})}=1\ .
\end{equation}
One may verify that for the parameters $Y_{n_2},\ldots,Z'_{n_3}$
in general position the system of equations \r{per23-sol} has
exactly $g=(N_2-1)(N_3-1)$ non-equivalent solutions (equivalence
means that if $(p,q)$ is a solution of (\ref{per23-sol}), then
$(q,p)$ is also the solution). Let us choose this set of solutions
as sequence \r{sequence} leaving parameters $f_k$, $k=1,\ldots,g$
to be free.

Using this freedom let us redefine the `amplitudes' $f_k$ as follows:
\begin{equation}\label{redefinition}
f_k=F_k\cdot\sigma_k(X_k),
\end{equation}
where we recall that parameters $p_k$ and $q_k$ of the
functions \r{S-fun} are already fixed by the system of equations
\r{per23-sol}, while parameters $X_k$ are still free. Let us consider
the solutions for the tau-functions $\tau_{2,\pot}$ and
$\tau_{3,\pot}$ given by the
Proposition~\ref{prop44} and
with redefined amplitudes \r{redefinition} at the value of the
discrete coordinate $n_1=0$. Using the freedom in parameters $X_k$
we send
\begin{equation}\label{send}
X_k\mapsto p_k\ .
\end{equation}
It is clear that in this limit all components of the set
$\{I_{\pot:\,k}\}_{k=0}^{g-1}$ vanish and according to
definition \r{casor} we will have $\tau_{2,\pot}=\tau_{3,\pot}=1$.
In other words, we obtained for these tau-functions at $n_1=0$
a zero-soliton solution. Let us repeat the same procedure at
$n_1=1$. Namely, again consider generic solution for these
tau-functions given by the Proposition~\ref{prop44} and then
make a limit \r{send}. It is clear that now in the set
$\{I_{\pot:\,k}\}_{k=0}^{g-1}$ one element which correspond to
$k=0$ will survive and as result we obtain one-soliton solution.
Increasing $n_1$ results in increasing the number of solitons.
One may see that the maximal number of solitons which can be
reached by this procedure is equal to $g=(N_2-1)(N_3-1)$.
See detailed description of this phenomena in the simplest situation
discussed in the paper \cite{RTC}.

But before conclude this section we would like to give one more
explanation of the solitons creation procedure. Let us consider
the equation \r{ttt} at $n_1=0$ for $\alpha,\beta,\gamma=1,2,3$
and homogeneous or zero-soliton $\tau$-functions
$\tau^{(0)}_{2:n_2,n_3}=\tau^{(0)}_{3:n_2,n_3}=1$. This is a
difference with respect to ``space'' coordinates $n_2$ and $n_3$
linear equation for the function $\tau_{1:n_2,n_3}$. Using simple
algebra one may verify that besides trivial solution
$\tau_{1:n_2,n_3}=1$ this equation has a solution of the form
$\tau_{1:n_2,n_3} = \prod_{m_2=0}^{n_2-1}
\frac{\sigma_0(Y'_{m_2})}{\sigma_0(Y_{m_2})} \prod_{m_3=0}^{n_3-1}
\frac{\sigma_0(Z'_{m_3})}{\sigma_0(Z_{m_3})}$, where parameter
$p_0$ of the function $\sigma_0$ is identified with $X_0$. The
complete solutions of linear difference equations is
\begin{equation}\label{step0}
\tau^{(1)}_{1:n_2,n_3}=1-F_0\ \prod_{m_2=0}^{n_2-1}
\frac{\sigma_0(Y'_{m_2})}{\sigma_0(Y_{m_2})}
\prod_{m_3=0}^{n_3-1}\frac{\sigma_0(Z'_{m_3})}{\sigma_0(Z_{m_3})}
\end{equation}
with arbitrary $F_0$, Now solving the equation \r{ttt} for
$\alpha,\beta,\gamma=2,3,1$ and $3,1,2$ with found
$\tau^{(1)}_{1:n_2,n_3}$ we can find the values of the
$\tau$-functions $\tau_{2:n_2,n_3}$ and $\tau_{3:n_2,n_3}$ at the
discrete time $n_1=1$. They will have similar to \r{step0}
one-soliton form. Using these solutions again in the equation
\r{ttt} with $\alpha,\beta,\gamma=1,2,3$ we will find two-soliton
solution for the function $\tau_{1:n_2,n_3}$ and then two-soliton
solutions for the functions $\tau_{2:n_2,n_3}$ and
$\tau_{3:n_2,n_3}$ at the next value of the discrete time $n_2=2$.
It is clear that this procedure can be continued and demonstrate
the equivalence of the discrete evolution parameter $n_1$ to the
number of solitons. This simple explanation finally justify that
investigated in this paper the discrete dynamic given by the
equations of motion \r{eq-of-mo1}--\r{eq-of-mo3} is a set of
consecutive B\"acklund transformations.

\section{Discussion}

As it is well known \cite{Sergeev}, the dynamics of parameters of 
3d spin models
is equivalent to the dynamics (\ref{eq-of-mo1}-\ref{eq-of-mo3}).
So the classical solutions discussed in this paper can be used for
the spin models of different types. The solutions with periodic
boundary conditions can be utilized for construction generic spin
model, such that their Boltzmann weights are parameterized by the
theta-functions of the higher genus. The integrability of such
spin models is based on the modified tetrahedron equation
\cite{GPS}. This should include the different generalization of
the famous Chiral Potts model.

On the other hand the solitonic solutions are convenient for
completely inhomogeneous ZBB model and open a way to develop the
quantum separation of variables for ZBB model \cite{svan}. In
particular, these solutions with parameters $f_k\neq 0$  allows
one to construct the complete family of isospectral deformations
of ZBB transfer-matrix (see also \cite{RTC} for realization of
this program in case of relativistic Toda chain model with spin
degrees of freedom).

\section{Acknowledgment}

This work was supported in part by the grants INTAS OPEN
00-00055 and CRDF RM1-2334-MO-02. S.P.'s work was supported by the
grant RFBR 01-01-00539, grant for support of scientific schools
RFBR 00-15-96557 and grant of Heisenberg-Landau Program
HLP-2002-11. S.S's work was supported by the grant RFBR
01-01-00201 and by Grant-in-Aid for scientific research of Japan
Society for the Promotion of Science.


\begin{thebibliography}{**}

\bibitem{Sergeev} S. Sergeev. A three-dimensional quantum integrable
mapping, \emph{Theoretical and Mathematical Physics}, \textbf{118}
(1999) 479-487; Quantum 2+1 evolution model, \emph{J. Phys. A:
Math. Gen.} \textbf{32} (1999) 5693-5714; Solitons in a 3d
integrable model, \emph{Phys. Lett.} \textbf{A 265} (2000)
364-368; Auxiliary transfer matrices for three dimensional
integrable models, \emph{Theoretical and Mathematical Physics},
\textbf{124} (2000) 1187-1201; Complex of three dimensional
integrable models, \emph{J. Phys. A: Math. Gen.} \textbf{34}
(2001) 10493-10503.

\bibitem{Kr78} I. Krichever. Algebraic curves and nonlinear difference
equations. \emph{Russian Math. Survey} \textbf{33} (1978)
215--216.

\bibitem{Hirota} R. Hirota. Discrete analogue of a generalized Toda equation'',
\emph{J. Phys. Soc. Jpn.} \textbf{50} (1981) 3785-3791.



\bibitem{Mumford} D. Mumford. Tata lectrures on Theta I,II.
Boston-Basel-Stuttgart, Birkh\"auser, 1983, 1985.

\bibitem{Fay} J.D. Fay. Theta Functions on Riemann Surfaces.
\emph{Lect. Notes in Math.}, \textbf{352} (1973).

\bibitem{Springer} G. Springer. Introduction to Riemann
Surfaces. Chelsea Pub. Co. New-York, sc. ed., 1981.

\bibitem{SNMP} S. Sergeev. On exact solution of a classical 3d
integrable model. \emph{Journal of Nonlinear Mathematical
Physics}, \textbf{27} (2000) 57-72.

\bibitem{RTC} S. Pakuliak, S. Sergeev. Quantum relativistic toda chain at root
of unity: isospectrality, modified $Q$-operators and functional
Bethe ansatz. Preprint MPI-2002-45 (nlin.SI/0205037).

\bibitem{svan} S. Sergeev. Functional equations and separation of variables
for 3d spin models, Preprint MPI-2002-46.

\bibitem{GPS} G. von Gehlen, S. Pakuliak, S. Sergeev. Explicit free
parametrization of the modified tetrahedron equation. In preparation.

\end{thebibliography}
\end{document}